\documentclass[preprint,aps,amssymb,showpacs,superscriptaddress,nofootinbib]{revtex4}
\usepackage{amsmath}
\usepackage{accents}
\usepackage{color}
\usepackage{graphicx}
\definecolor{DarkRed}{rgb}{0.545098,0.000000,0.000000}
\definecolor{blue1}{rgb}{0.000000,0.000000,1.000000}
\definecolor{red1}{rgb}{1.000000,0.000000,0.000000}
\definecolor{indianred}{rgb}{0.803922,0.360784,0.360784}

\newcommand{\Case}[2]{{\textstyle \frac{#1}{#2}}}

\newcommand{\U}[1]{\underline{#1}}

\begin{document}
%
\today \hfill IMSc/2016/12/07

\title{Cosmological Horizon and the Quadrupole Formula in de Sitter
Background}

\author{Ghanashyam Date}
\email{shyam@imsc.res.in}
\affiliation{The Institute of Mathematical Sciences, HBNI, \\
CIT Campus, Chennai-600 113, INDIA.}

\author{Sk Jahanur Hoque}
\email{jahanur@imsc.res.in}
\affiliation{The Institute of Mathematical Sciences, HBNI, \\
CIT Campus, Chennai-600 113, INDIA.}

\begin{abstract} 
An important class of observables for gravitational waves consists of
the fluxes of energy, momentum and angular momentum carried away by them
and are well understood for weak gravitational waves in Minkowski
background. In de Sitter background, the future null infinity,
$\mathcal{J}^+$, is space-like which makes the meaning of these
observables subtle. A spatially compact source in de Sitter background
also provides a distinguished null hypersurface, its {\em cosmological
horizon}, $\mathcal{H}^+$. For sources supporting the short wavelength
approximation, we adopt the Isaacson prescription to define an effective
gravitational stress tensor. We show that the fluxes computed using this
effective stress tensor can be evaluated at $\mathcal{H}^+$, match with
those computed at $\mathcal{J}^+$ and also match with those given by
Ashtekar et al at $\mathcal{J}^+$ {{\em at a coarse grained
level}}.
\end{abstract}


\maketitle

\section{Introduction}

Weak gravitational field of a spatially compact source is identified as
a perturbation about a background space-time which is a solution of the
Einstein equation in the source free region. In the presence of a
positive cosmological constant, the background space-time is the de
Sitter space-time. Unlike the Minkowski background for the zero
cosmological constant, de Sitter space-time has different patches eg the
global patch ($\mathbb{R}\times S^3$), a Poincare patch and a static
patch. In the cosmological context, a Poincare patch is appropriate
which is what we focus on.  A solution at the {\em linearized level},
valid throughout the Poincare patch and extending to the future null
infinity ${\cal J}^+$, is available in \cite{deVega, ABKIII, DateHoque}.
However the space-like character of the $\mathcal{J}^+$ poses challenges
for defining energy, momentum and their fluxes. 

Let us recall that the cleanest articulation of `infinity' arises in the
conformal completion of physical space-times. Conformal completion preserves
the light cone structure of the physical space-time and naturally identifies
boundary components, ${\cal J}^{\pm}$ where time-like and null geodesics
`terminate'. The causal nature of these boundary components is determined by
the asymptotic form of `source-free' equations: ${\cal J}^{\pm}$ are null when
$\Lambda = 0$ and space-like for $\Lambda > 0$ (time-like for $\Lambda < 0$).
These boundary components serve to define out-going (in-coming) fields as
those solutions of the asymptotic equations that have suitably finite limiting
values on ${\cal J}^+ ({\cal J}^-)$. It is then a result that the Weyl tensor
of out-going fields evaluated along out-going null geodesics, has a definite
pattern of fall-off in inverse powers of an affine parameter along the
geodesics (the peeling-off theorem) \cite{Sachs61, PenroseZeroMass}. This
enables one to identify the leading term as representing gravitational
radiation (far field of a source), in a coordinate invariant manner. It is
conveniently described in terms of the Weyl scalars which are defined with
respect to a suitable null tetrad.  When ${\cal J}^+$ is null, a null tetrad
at a point $p \in \mathcal{J}^+$ is uniquely determined (modulo real scaling
and rotation) by the tangent vector $\ell^{\mu}$ of an outgoing null geodesic
reaching $p$, {\em and } the null normal $n^{\mu}$, satisfying $\ell\cdot n =
-1$.  Clearly as the null geodesic changes its direction, $\ell$ changes but
not $n$ and hence the Weyl scalar $\Psi_4 \ (:=
C_{\mu\nu\rho\sigma}n^{\mu}\bar{m}^{\nu}n^{\rho}\bar{m}^{\sigma})$ remains
unchanged. Its non-zero value can be taken as showing the presence of
gravitational radiation. This feature is lost when the ${\cal J}^+$ is
space-like. Now the null vector $n^{\mu}$, with $\ell\cdot n = -1$, is chosen
to be in the plane defined by $\ell^{\mu}$ and the (time-like) normal
$N^{\mu}$.  Clearly, as $\ell^{\mu}$ changes, so does $n^{\mu}$ and {\em none}
of the Weyl scalars is invariant. An invariant characterization of
gravitational radiation is no longer available \cite{PenroseZeroMass}.

The de Sitter space-time also has the so called observer horizons - boundary
of the causal past of an observer's end point on $\mathcal{J}^+$.  In
particular, for a spatially extended but compact source, the worldlines of
different components of the source, must reach the same point on
$\mathcal{J}^+$ to maintain a {\em finite} physical separation among them. A
spatially compact source then defines (its) {\em cosmological horizon} as the
past light cone of the common point on $\mathcal{J}^+$ where the source world
tube converges. Equally well, {\em any observer} who remains at a finite
physical distance from the compact source for all times, must necessarily lie
within the cosmological horizon i.e.  within the static patch bounded by the
cosmological horizon. Unlike the $\mathcal{J}^+$, the cosmological horizon is
a null hypersurface but shares with $\mathcal{J}^+$ the property, that
whichever curve meets a point on it, can never causally intersect the world
tube of the spatially compact source.  In other words, once any
energy/momentum/angular momentum is carried away across the cosmological
horizon, it is `lost' from the source forever.  We would like to explore to
what extent and under what conditions may we regard the cosmological horizon
as a ``substitute'' for the {\em future null infinity}.

It is obvious at the outset that the out-going null geodesics emanating from
the source intersect the cosmological horizon at a {\em finite} value of any
affine parameter and it can be chosen to be 1 by a suitable normalization.
Such a normalized affine parameter equals the ratio of the physical distance
from the source to ${\sqrt{3/\Lambda}} \sim 10^{26} m \sim 10 Gpc$. All
spatially compact sources may be taken to lie within a sphere of radius $\sim
\Lambda^{-1/2}$. Furthermore, only sources varying over cosmological time
scales, will have comparable wavelengths.  Thus, most sources producing
gravitational waves would have wavelengths far smaller than $\Lambda^{-1/2}$
and any wave crossing the horizon may be taken to be a `far zone field'.
Cosmological horizon being a null hypersurface, a $\Psi_4$ can be defined on
it, independent of the  null geodesics meeting the horizon.  A notion of
radiation based on asymptotic behaviour of fields is physically useful,
provided there are suitable definitions of fluxes of energy-momentum, and
angular momentum in terms of these asymptotic fields. And there are many such
definitions.

One of the definition of such conserved quantities is based on the
covariant phase space framework \cite{AshtekarBombelliReula, ABKII}. In
the context of the linearised theory, it exploits the phase space
structure of the space of solutions and defines a manifestly gauge
invariant and conserved `Hamiltonian' corresponding to each of the seven
{\em isometries} of the Poincare patch.  Although defined on each
space-like hypersurface of the Poincare patch, the simplest expressions
result for evaluation at ${\cal J}^+$. Thus, the conserved quantities
are directly expressed in terms of the asymptotic fields. 

For sources which are sufficiently rapidly varying (relative to the scale set
by the cosmological constant), there is an alternative identification of
gravitational waves as {\em ripples on a background} within the so called {\em
short wave approximation} \cite{Isaacson, MTW}.  Furthermore, it is possible
to define an {\em effective gravitational stress tensor}, $t_{\mu\nu}$ for the
ripples. For vanishing $\Lambda$, it is symmetric, conserved and gauge
invariant. For non-zero $\Lambda$ it is {\em not} gauge invariant but the
gauge violations are suppressed by powers of $\sqrt{\Lambda}$. It is very
convenient to have such a stress tensor to define and compute fluxes of energy
and momenta carried by the ripples across {\em any} hypersurface. 

We use the fluxes defined using the effective gravitational stress tensor and
show that for the retarded solution given in \cite{deVega, ABKIII, DateHoque},
the fluxes of energy and momentum across the cosmological horizon exactly
equal the corresponding fluxes across the $\mathcal{J}^+$. Furthermore, these
fluxes computed at $\mathcal{J}^+$ also equal the fluxes defined in the
covariant phase space framework, \cite{ABKIII} albeit {at a coarse grained
level (See equation (\ref{CoarseGrained}))}.  The instantaneous power received
at infinity matches with that crossing the horizon. This is our main result.

The paper is organised as follow.

In section \ref{BACKGROUND}, we summarise various details needed to establish
our result. Most are available in the cited literature and are collected here
for self contained reading. It is divided in three subsection. In the
subsection \ref{SOLUTION}, we recall the solution at the linearised level
\cite{deVega, ABKIII, DateHoque} for which the fluxes will be evaluated.  We
specify and denote the (spatial components of) the {\em exact retarded}
solution by $\mathcal{X}_{ij}$.  This is approximated when the source
dimension is much smaller than the distance to the source. The leading term is
the {\em approximated retarded} solution and is denoted by $\chi_{ij}$.
Physical solutions have to satisfy the gauge conditions imposed in simplifying
the linearised equation. This is achieved by extracting the (spatial) {\em
transverse and traceless} (TT) part of the solution which is denoted by
$\mathcal{X}^{TT}_{ij}$.  For the approximated solution, the TT part is
conveniently extracted by an algebraic projection to the same level of
approximation. The {\em algebraically projected transverse, traceless} part of
the approximated solution is denoted by $\chi_{ij}^{tt}$ and used throughout.
We also collect relevant expressions for subsequent use. A table of notation
is included at the end of this subsection. 
In subsection \ref{COVARIANT} we summarise the covariant phase space framework
and recall the definitions of the fluxes and quadrupole power from
\cite{ABKIII}. The energy momentum fluxed defined here are compared to those
defined in the next subsection.
In subsection \ref{EFFECTIVEStressTensor}, we discuss the Isaacson
prescription {\em adapted} to the presence of the cosmological constant and
present the definition of the {\em ripple tensor} in eq.
(\ref{RippleTensorFinal}) which is used in the next section. 

Section \ref{ENERGY} is divided into three subsections.  In the subsection
\ref{FLUXComputations}, we present computations of the energy flux for the
$\chi_{ij}^{tt}$ across various hypersurfaces.  In particular we show that the
fluxes across the out-going null hypersurfaces are zero, implying for example,
that the energy propagation is sharp. Subsection \ref{MOMENTUMFluxes} contains
the fluxes for the momentum and the angular momentum. In the subsection
\ref{FROMtttoTT} we discuss how the computations can be extended to
$\chi^{TT}_{ij} := (\mathcal{X}_{ij}^{TT})_{approx}$. 

In section \ref{COSMOLOGICALHorizon}, we discuss applications of these flux
computations and establish our main results.  The final section
\ref{FINALSection} concludes with a discussion. An appendix is included to
illustrate an averaging procedure.

\vskip 0.3cm

\section{Preliminaries} \label{BACKGROUND}
In this section we summarise and assemble already available relevant details
needed for our main result, with the main citations included in the subsection
headings. 

\subsection{Weak gravitational field of interest \cite{deVega, ABKIII,
DateHoque}} \label{SOLUTION}
Weak gravitational fields are understood as perturbations about a background
specified in the form, $g_{\mu\nu} := \bar{g}_{\mu\nu} + \epsilon h_{\mu\nu}$.
The background $\bar{g}_{\mu\nu}$ is chosen to be a solution of the source
free Einstein equation with a positive cosmological constant. The Einstein
equation for $g_{\mu\nu}$, expanded to first order in $\epsilon$, gives the
linearised Einstein equation for $h_{\mu\nu}$. The {\em physical
perturbations} are understood as the equivalence classes of solutions
$h_{\mu\nu}$, with respect to the {\em gauge transformations}: $\delta
h_{\mu\nu}(x) = {\cal L}_{\xi}\bar{g}_{\mu\nu}(x) =
\bar{\nabla}_{\mu}\xi_{\nu} + \bar{\nabla}_{\nu}\xi_{\mu}$. In terms of the
trace reversed combination $\tilde{h}_{\mu\nu} := h_{\mu\nu} -
\Case{1}{2}\bar{h}_{\mu\nu}(\bar{g}^{\alpha\beta}h_{\alpha\beta})$, the
linearised equation takes the form,
\begin{equation}
\frac{1}{2}\left[ - \bar{\Box} \tilde{h}_{\mu\nu} + \left\{
\bar{\nabla}_{\mu}B_{\nu} + \bar{\nabla}_{\nu}B_{\mu} -
\bar{g}_{\mu\nu}(\bar{\nabla}^{\alpha}B_{\alpha})\right\}\right] +
\frac{\Lambda}{3}\left[\tilde{h}_{\mu\nu} - \tilde{h}\bar{g}_{\mu\nu}\right] ~
= ~ 8\pi T_{\mu\nu} \label{LinEqn}
\end{equation}
where, $B_{\mu} := \bar{\nabla}_{\alpha}\tilde{h}^{\alpha}_{~\mu}$.  The gauge
freedom is exploited subsequently to simplify the equation. 

\begin{figure}[htb]
\includegraphics[width=0.5\textwidth]{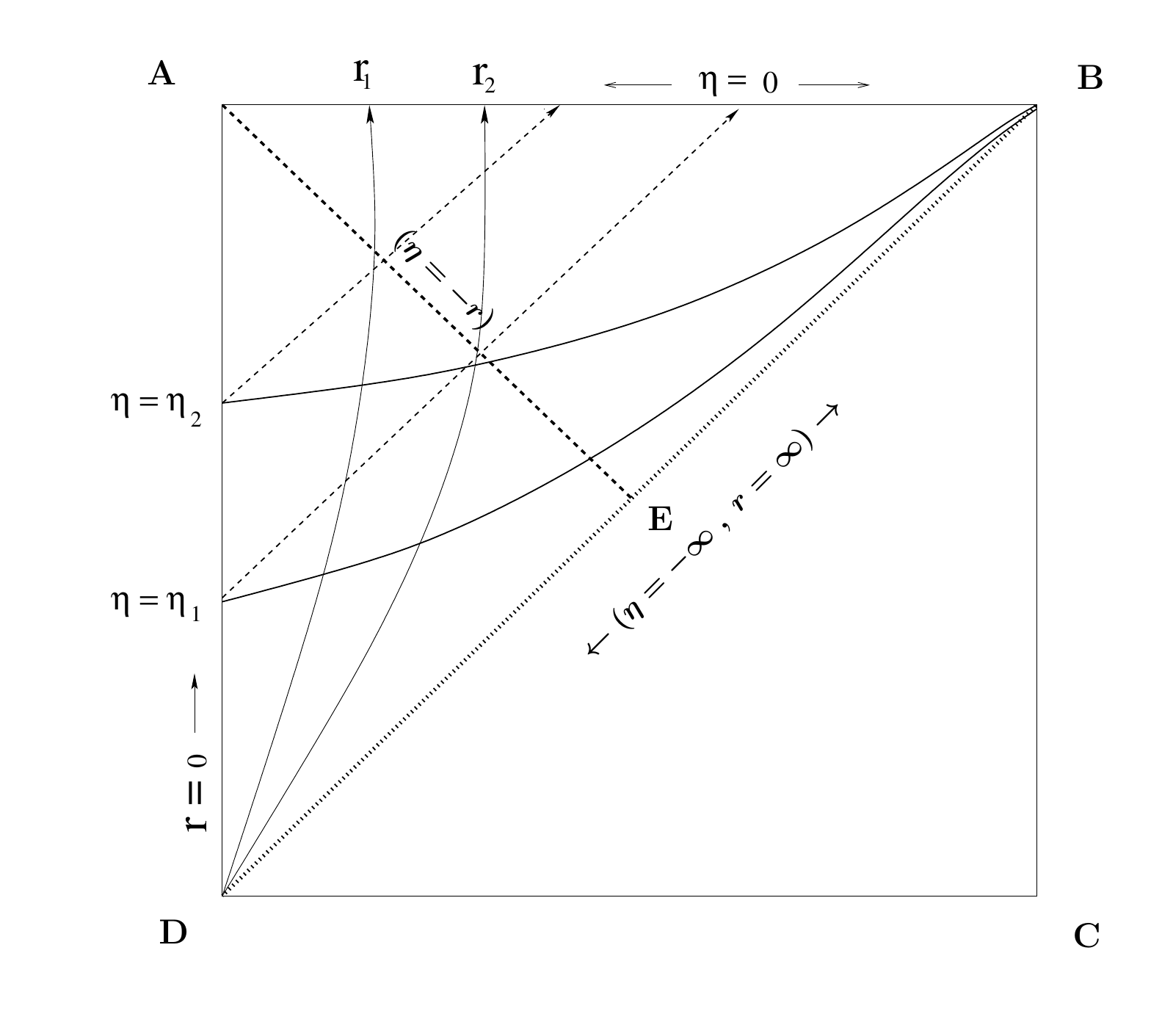}
\caption{The full square is the Penrose diagram of de Sitter space-time
with generic point representing a 2-sphere. World tube of a spatially
compact source is taken to be centred on the line DA. The corresponding
Poincare patch is labeled ABD and is covered by the Poincare chart
$(\eta, r, \theta, \phi)$.  The line BD does not belong to the chart.
The line AB denotes the {\em future null infinity}, $\mathcal{J}^+$
while the line AE denotes the {\em cosmological horizon},
$\mathcal{H}^+$ of the source.  The region AED is the {\em static patch}
admitting the stationary Killing vector, $T$ of eqn.
(\ref{StationaryKilling}). Two constant $\eta$ space-like hypersurfaces
are shown with $\eta_2 > \eta_1$. The two constant $r$, time-like
hypersurfaces have $r_2 > r_1$ while the two dotted lines at 45 degrees,
denote the out-going null hypersurfaces emanating from $\eta = \eta_1,
\eta_2$ on the world line at $r = 0$. }
\label{DeSitterPenrose}
\end{figure}

In the present context, the background space-time is taken to be the
{\em Poincare patch} of the de Sitter space-time (see figure
\ref{DeSitterPenrose}) which admits a conformally flat form of the
background metric in coordinates ($\eta, x^i)$~,
\begin{eqnarray}
ds^2 & = & \frac{1}{H^2\eta^2}\left[ - d\eta^2 + \sum_i (dx^i)^2
\right]\ , ~~ \eta \in ( - \infty , 0 )~~,~~ x^i \in \mathbb{R} ~~,~~ H
:= \sqrt{\frac{\Lambda}{3}} . \label{ConformallyFlat}
\end{eqnarray}
The future null infinity is approached as $\eta \to 0_-$ while the
$\eta \to -~\infty$ corresponds to the FLRW singularity. The conformal
factor is $a^2(\eta) := (H\eta)^{-2}$.

The linearised equation is simplified by imposing the {\em generalised
transverse gauge} conditions: $B_{\mu} =
\Case{2\Lambda}{3}\eta\tilde{h}_{0\mu}$ The conformal factor can be
scaled out by using the fields $\tilde{\chi}_{\mu\nu} :=
a^{-2}\tilde{h}_{\mu\nu}$ and the linearised equation (with source
included) then takes the form\footnote{From now on in this subsection,
the tensor indices are raised/lowered with the Minkowski metric.}
\cite{deVega},
\begin{eqnarray}
-~16 \pi T_{\mu\nu} & = & \Box \tilde{\chi}_{\mu\nu} +
\frac{2}{\eta}\partial_0\tilde{\chi}_{\mu\nu} -
\frac{2}{\eta^2}\left(\delta_{\mu}^0\delta_{\nu}^0\tilde{\chi}_{\alpha}^{~\alpha}
+ \delta_{\mu}^0\tilde{\chi}_{0\nu} + \delta_{\nu}^0\tilde{\chi}_{0\mu}
\right) \ .  \label{LinEqnChi} \\ 
0 & = & \partial^{\alpha}\tilde{\chi}_{\alpha\mu} +
\frac{1}{\eta}\left(2 \tilde{\chi}_{0\mu} + \delta_{\mu}^0
\tilde{\chi}_{\alpha}^{~\alpha} \right) \hspace{2.5cm} (\mbox{gauge
condition}). \label{ChiGauge}
\end{eqnarray}

It is further shown in \cite{deVega} that the residual gauge invariance
is exhausted by imposing the additional gauge conditions:
$\tilde{\chi}_{0\alpha} = 0 = \hat{\tilde{\chi}} (:= \tilde{\chi}_{00} +
\tilde{\chi}_{i}^{~i})$. The gauge condition (\ref{ChiGauge}) then
implies that {\em physical perturbations} may be characterised solutions
of (\ref{LinEqnChi}) which satisfy the spatial transverse, traceless
condition, or {\em spatial TT} for short: $\partial^{j}\tilde{\chi}_{ji}
= 0 = \tilde{\chi}^{k}_{~k}$. Thanks to the decoupled equations, it
suffices to focus on the spatial components of the equation.

{The {\em exact retarded solution}} is given by,
\begin{eqnarray}
{\mathcal{X}_{ij}(\eta, x)} & = & 4 \int d^3x' \frac{\eta}{|x -
x'|(\eta
- |x - x'|)} \left.  T_{ij}(\eta', x') \right|_{\eta' = \eta - |x -
  x'|} \nonumber \\
& & \hspace{0.5cm} 
+ ~ 4 \int d^3x' \int_{- \infty}^{\eta - |x
-x'|}d\eta'\frac{T_{ij}(\eta', x')}{\eta'^2} \label{RadiativeSoln} 
\end{eqnarray}
The spatial integration is over the matter source confined to a
compact region and is finite. {The second term in the eqns.
(\ref{RadiativeSoln}) is the {\em tail term}.} This particular
solution does {\em not} satisfy the spatial TT conditions. Using the
transverse, traceless decomposition of the tensor fields, the TT part,
$\mathcal{X}_{ij}^{TT}$ is extracted which represents the {\em
physical retarded field} due to the source. 

For $|\vec{x}| \gg |\vec{x}'|$, we can approximate $|\vec{x} - \vec{x}'|
\approx r := |\vec{x}|$. This allows us separate out the $\vec{x}'$ dependence
from the $\eta - |x - x'|$.  The so {\em approximated retarded solution},
$\chi_{ij}$, is given by,
{ \begin{eqnarray}
\mathcal{X}_{ij} & = & \chi_{ij}(\eta, x) + o(r^{-1}) \hspace{3.0cm}, ~~~
\mbox{with} \nonumber \\
\chi_{ij}(\eta, x) & :=  & \left. 4 \frac{\eta}{r(\eta - r)} \int d^3x'
T_{ij}(\eta', x') \right|_{\eta' = \eta - r}  +  4 \int_{- \infty}^{\eta -~r
}d\eta'\frac{1}{\eta'^2}\int d^3x' T_{ij}(\eta', x') \label{RadiativeSoln1}
\end{eqnarray} }
{We will work with the approximated solution. Note that $\chi_{ij}$ depends on
$\vec{x}$ only through $r = |\vec{x}|$.} The spatial integral of $T_{ij}$ can
be simplified using moments.  This is done through the matter conservation
equation. 

To define these moments, introduce the orthonormal tetrad
$f^{\alpha}_{~\U{a}} := - H\eta\delta^{\alpha}_{~\U{a}}$ and denote the
frame components of the source stress tensor as: $\rho :=
H^2\eta^2T_{\U{00}}, ~ \pi := H^2\eta^2T_{\U{ij}}\delta^{\U{ij}}$.
Define moment variable $\bar{x}^{\U{i}} = f^{\U{i}}_{~\alpha}x^{\alpha}
= - (\eta H)^{-1}\delta^{\U{i}}_{~j}x^j := a(t)x^{\U{i}}$. Two sets of
moments are defined as functions of $\eta$, (or of $t$ defined through
$\eta := - H^{-1}e^{-Ht}$) as,
\begin{eqnarray}\label{PhysicalMoments}
{Q}^{\U{ij}}(t) ~ := ~ \int_{Source(t)} d^3x \ a^{3}(t)\rho(t,\vec{x})
\bar{x}^{\U{i}}\bar{x}^{\U{j}} & ~,~ &
\bar{Q}^{\U{ij}}(t) ~ := ~ \int_{Source(t)} d^3x \
a^{3}(t)\pi(t,\vec{x}) \bar{x}^{\U{i}}\bar{x}^{\U{j}} \ .
\end{eqnarray}

In terms of these, the {approximated} retarded solution is given by,
\begin{eqnarray} \label{CosmoGWAmplitude} 
\chi_{ij}(\eta, \vec{x}) & = & \frac{1}{r}f_{ij}(\eta_{ret}) +
g_{ij}(\eta_{ret}) + \hat{g}_{ij}\hspace{5.2cm} \mbox{with,} \\
f_{ij}(\eta_{ret}) & := & \frac{2}{a(\eta_{ret})}\left[{\cal L}^2_T{Q}_{ij} +
2H{\cal L}_T{Q}_{ij} + H{\cal L}_T{\bar{Q}}_{ij} + 2H^2 \bar{Q}_{ij}\right] ~,
\\
g_{ij}(\eta_{ret}) & := & - 2H\left[ {\cal L}^2_T{Q}_{ij} + H{\cal
L}_T{Q}_{ij} + H{\cal L}_T{\bar{Q}}_{ij} + H^2 \bar{Q}_{ij}\right] ~
\hspace{1.5cm}\mbox{and,} \\
\hat{g}_{ij} & := & \left. -2H^2\left[\mathcal{L}Q_{ij} +
H\bar{Q}_{ij}\right]\right|_{-\infty}
\end{eqnarray}
All moments are evaluated at the retarded $\eta_{ret} := (\eta - r)$,
$a(\eta_{ret}) := - (H\eta_{ret})^{-1}$ and ${\cal L}_T$ denotes the Lie
derivative with respect to the time translation Killing vector defined in
equation (\ref{StationaryKilling}) below. On the moments, it is given by,
\begin{equation} \label{LieDerivative}
{\cal L}_{T}Q_{ij} ~ = -H(\eta\partial_{\eta}+r\partial_{r})Q_{ij}-2HQ_{ij}
=-H(\eta-r)\partial_{\eta}Q_{ij}-2HQ_{ij}
=\partial_{t}Q_{ij}|_{t_{ret}}-2HQ_{ij} 
\end{equation} 

In equation \eqref{CosmoGWAmplitude}, the first term is the contribution
of the so called sharp term while the second and the third terms denote
the tail contributions. The tail contribution has separated into a term
which depends on retarded time, $(\eta -r) $ only, just as the sharp
term does, and the contribution from the history of the source is given
by the limiting value at $\eta=-\infty$.  This expression is valid {\em
as the leading term for $|\vec{x}| \gg |\vec{x}'|$}. There is no TT
label on these expressions. While the solution has tail term, it will
turn out that the energy propagation is sharp.

For future use in section \ref{FLUXComputations}, we display the
derivatives of $\chi_{ij}$. Since $\chi_{ij}$ depends on $\vec{x}$ only
through $r$, we need only the derivatives with respect to $\eta$ and
$r$.  On functions of $\eta_{ret}$, $\partial_{r} = - \partial_{\eta}$
and we can replace the r-derivatives in favour of $\eta-$derivatives.
Hence, 
\begin{equation}\label{ChiDerivatives}
\partial_{\eta}\chi_{ij} ~ = ~ \frac{1}{r}\partial_{\eta} f_{ij} +
\partial_{\eta} g_{ij}  ~~~~,~~~~ 
\partial_r \chi_{ij}(\eta,r) ~ = ~ 
- \partial_{\eta}\chi_{ij}  -\frac{f_{ij}}{r^2} \ .
\end{equation}

{There is a well known {\em algebraic projection} method to
construct spatial tensors which satisfy the spatial TT condition {\em to
the leading order in $r^{-1}$}. Since the approximated solution is also
valid to $o(r^{-1})$, we may use this convenient method.

For the unit vectors $\hat{x}$ denoting directions, define the
projectors
\begin{equation}\label{TTProjection}
P_i^{~j} ~ := ~ \delta_i^{~j} - \hat{x}_i\hat{x}^{j} ~~,~~ 
\Lambda_{ij}^{~~kl} ~ := ~ \frac{1}{2}(P_i^{~k}P_j^{~l} +
P_i^{~l}P_j^{~k} - P_{ij}P^{kl})~~,~~ \chi_{ij}^{tt} ~ := ~
\Lambda_{ij}^{~~kl}\chi_{kl} 
\end{equation}
We have used the notation of `tt' to refer to the algebraically
projected transverse, traceless part as in \cite{ABKIII}.  Noting that
on $\chi_{ij}$ the spatial derivative is $\partial^j =
\hat{x}^j\partial_r$, it follows that, 
\begin{eqnarray}
\partial_{\eta}(\chi_{ij}^{tt}) & = & (\partial_{\eta}\chi_{ij})^{tt} ~
, ~ \partial_{r}(\chi_{ij}^{tt}) = (\partial_{r}\chi_{ij})^{tt} ~~,~~ \\
\partial_m(\chi_{ij}^{tt}) & = &
(\partial_{m}\Lambda_{ij}^{~~kl})\chi_{kl} +
\hat{x}_m(\partial_r\chi_{ij})^{tt} \\
\therefore \partial^j(\chi_{ij}^{tt}) & = &
\hat{x}^j\Lambda_{ij}^{~~kl}\partial_r\chi_{ij} +
(\partial^{j}\Lambda_{ij}^{~~kl})\chi_{kl} ~ = ~ 0 + o(r^{-1}) \ ;
\hspace{0.5cm}\mbox{where we used,} \label{Transversality} \\
\partial_{m}\Lambda_{ij}^{~~kl} & = & -
\frac{1}{r}\left[\hat{x}_i\Lambda_{mj}^{~~~kl} +
\hat{x}_j\Lambda_{mi}^{~~~kl} + \hat{x}^k\Lambda_{ijm}^{~~~~l} +
\hat{x}^l\Lambda_{ijm}^{~~~~k}\right] ~ = ~ o(r^{-1}) \ . \label{ProjectorDerivative} 
\end{eqnarray}
The tracelessness of $\chi_{ij}^{tt}$ is manifest and hence
$\chi_{ij}^{tt}$ satisfies the spatial TT condition to $o(r^{-1})$.
}

Using the derivatives of $\chi_{ij}$ given in \eqref{ChiDerivatives}, we
can write (the right hand sides denote row vectors of the $\mu = \eta$
and $\mu = m$ components),
\begin{eqnarray}\label{DelMuChi}
\partial_{\mu}\chi_{ij}^{tt} & = & (\partial_{\eta}\chi_{ij}^{tt}~,~
\hat{x}_m\partial_r\chi_{ij}^{tt} +
(\partial_{m}\Lambda_{ij}^{~~kl})\chi_{kl}) \\
& = & (\partial_{\eta}\chi_{ij}^{tt})(1~,~ -\hat{x}_m) -
\left(\frac{f_{ij}^{tt}}{r^2}\right)(0~, ~\hat{x}_m) \nonumber \\
& &  \hspace{1.0cm} -\frac{1}{r}(0~,~
\left[\hat{x}_i\Lambda_{mj}^{~~~kl} + \hat{x}_j\Lambda_{mi}^{~~~kl} +
\hat{x}^k\Lambda_{ijm}^{~~~~l} +
\hat{x}^l\Lambda_{ijm}^{~~~~k}\right]\chi_{kl}) .
\end{eqnarray}
The first term is proportional to a null vector. The second term is
proportional to the space-like, radial vector. The third is again a
space-like vector. Both the second and the third terms are down by a
power of $r$ relative to $\chi_{ij}$ and therefore also relative to the
first term.  We will see later in the calculation of the fluxes that for
energy and momentum, the second and the third terms can be neglected.
However for flux of angular momentum, the third term is crucial.  {\em
When} the second and the third terms can be neglected, the effective
gravitational stress tensor turns out to correspond to an {\em out-going
null dust} with energy density proportional to
$\langle\partial_{\eta}\chi_{mn} \partial_{\eta}\chi^{mn}\rangle$. 

Finally, we note the isometries of the Poincare patch.  There are seven
globally defined Killing vectors on the Poincare chart, corresponding to
energy, 3 momenta and 3 angular momenta \cite{ABKI, ABKIII}.  They are
given by (up to constant scaling):
\begin{eqnarray} 
\mbox{Generator of time translation} & : & T = -H(\eta\partial_{\eta} +
x^i\partial_i) \label{StationaryKilling}\\
\mbox{Generators of space translation} & : & S_{(i)} = \partial_i
\label{SpaceKilling}\\
\mbox{Generators of space rotations} & : & L_{(j)} =
\epsilon_{jk}^{~~i}x^k\partial_i \ \label{RotationKilling}.
\end{eqnarray}
We focus on the {\em time translation vector field}, which is time-like
in the static patch, null on the cosmological horizon and and space-like
beyond it. In particular it is space-like and tangential to ${\cal
J}^+$.

Many different symbols used for the retarded solution, its
approximations, their TT parts and different `radiation fields' are
summarised below. 
\begin{center}
\begin{tabular}{ccll}
$\tilde{\chi}_{ij}$ & : & \hspace{0.3cm} generic solution of linearised
eqn. & \\
$\mathcal{X}_{ij}$ & : & \hspace{0.3cm} exact, retarded solution &
$~~~~$ eqn.  (\ref{RadiativeSoln}) \\
$ \mathcal{X}^{TT}_{ij}$ & : & \hspace{0.3cm} TT part of exact, retarded
solution & $~~~~$ eqn. (\ref{RadiativeSolnTT}) \\
$ \chi_{ij}$ & : & \hspace{0.3cm} approximated, retarded solution &
$~~~~$ eqn.  (\ref{RadiativeSoln1}) \\
$ \chi^{tt}_{ij}$ & : & \hspace{0.3cm} $\Lambda-$projection of
approximated, retarded solution & $~~~~$ eqn. (\ref{TTProjection}) \\
$\chi_{ij}^{TT}$ & : & \hspace{0.3cm} approximation of
$\mathcal{X}^{TT}_{ij}$ for $|\vec{x}'| \ll |\vec{x}|$ & $~~~~$ Above
eqn. (\ref{RTTDefn}) \\
& : & {\hspace{0.3cm} $\chi_{ij}^{TT}$ does {\em not} denote TT part of
$\chi_{ij}$}  & $~~~~$ {See footnote \ref{TTApproxCommute}.} \\
$\mathcal{R}_{ij}^{TT}$ & : & \hspace{0.3cm} Radiation field defined in
\cite{ABKIII} & $~~~~$ eqns. (\ref{EnergyAtInfinity},
\ref{RadiationField}) \\
$\mathcal{Q}_{ij}^{tt}$ & : & \hspace{0.3cm} Radiation field used
throughout & $~~~~$ eqns. (\ref{RttDefn}, \ref{RadnField}) \\
$\mathcal{Q}_{ij}^{TT}$ & : & \hspace{0.3cm} defined to equal
$2\partial_{\eta}\mathcal{M}_{ij}^{TT}$ & $~~~~$ eqn. (\ref{RTTDefn}) 
\end{tabular}
\end{center}

\subsection{Covariant phase space framework \cite{AshtekarBombelliReula,
ABKII, ABKIII}} \label{COVARIANT} 
Traditionally, the conserved energy, momentum etc are defined through
pseudo-tensors which have their shortcoming of not being covariant. The
framework of covariant phase space provides manifestly gauge invariant
definitions of the conserved quantities and is briefly recalled below.

Consider the space ${\cal C}$ of a class of solutions of the full
Einstein equation, satisfying stipulated boundary condition. At each
point of this space, the linearised solutions provide tangent vectors.
Under certain conditions, it is possible to define a {\em pre-symplectic
form} on the tangent spaces.  Every infinitesimal diffeomorphism of the
space-time, with suitable asymptotic behaviour, induces a vector field
on ${\cal C}$. Some of these lie in the kernel of the pre-symplectic
form and constitute `gauge directions' while the remaining ones
constitute (asymptotic) symmetries shared by the stipulated class of
solutions.  Modding out by the gauge directions (null space of the
pre-symplectic form), one imparts a symplectic structure to the space of
solutions, now denoted as $\Gamma \sim {\cal C}/gauge$. Under favourable
conditions, the vector fields on $\mathcal{C}$ corresponding to the
asymptotic symmetries descend to $\Gamma$ and generate infinitesimal
{\em canonical transformations}.  Their generating functions, or
`Hamiltonians', are candidates for representing energy, momenta, angular
momenta etc \cite{AshtekarBombelliReula}.

In \cite{ABKII, ABKIII}, this strategy is applied to the space of fully gauge
fixed solutions of the linearised equation and we summarise it below.
Isometries of the background, leave the covariant phase space itself invariant
and constitute canonical transformations. In the present context, the
Hamiltonians corresponding to the 7 isometries are the proposed definitions of
energy, linear momentum and angular momentum.

Explicitly, ${\cal C}$ denotes the solutions of the equation
(\ref{GaugeFixedLinearised}) together with the gauge fixing conditions
(\ref{GaugeCondns}): 
\begin{equation} \label{GaugeFixedLinearised}
\Box\tilde{\chi}_{ij} + \frac{2}{\eta}\partial_{\eta}\tilde{\chi}_{ij} = 0 
\end{equation}
\begin{equation} \label{GaugeCondns}
\partial^i\tilde{\chi}_{ij} = 0 ~~,~~ \tilde{\chi}_{ij}\delta^{ij} = 0
\end{equation}

A symplectic form is defined by an integral over a cosmological slice
$\Sigma_{\eta}$. A definition which has a smooth limit to
$\mathcal{J}^+$ ($\eta \to 0_+)$ is defined in terms of the {\em
electric part} of the perturbed Weyl tensor, ${\cal E}_{ij} := -(H
\eta)^{-1}[^{(1)} C^0_{~j0i}] = \Case{1}{ 2 H \eta^{2}}(\partial_{\eta}
\tilde{\chi}_{ij} + \eta~\nabla^{2}\tilde{\chi}_{ij}) = \Case{1}{ 2 H
\eta}(\partial^2_{\eta} -
\Case{1}{\eta}\partial_{\eta})\tilde{\chi}_{ij}$.  For two elements
$\tilde{\chi}, \U{\tilde{\chi}} \in {\cal C}$, the symplectic form is
defined by\cite{ABKIII},
\begin{equation}
\omega(\tilde{\chi},\U{\tilde{\chi}})= \frac{1}{16 \pi H}
\int_{\Sigma_{\eta}}d^{3}x (\tilde{\chi}_{ij} \U{{\cal E}}_{kl}-
\U{\tilde{\chi}}_{kl} {\cal E}_{ij})~ \delta^{ik} \delta^{jl}
\end{equation}
The $TT$ label on the $\tilde{\chi}$'s is suppressed.

A Killing vector $K$ of the de Sitter background, defines a vector field
$h^{(K)}_{ij} := {\cal L}_K h_{ij} = a^2({\cal L}_K\tilde{\chi}_{ij} + 2
(a^{-1}{\cal L}_K a) \tilde{\chi}_{ij}) =: a^2\tilde{\chi}^{(K)}_{ij}$, on the
space $\mathcal{C}$. This vector field generates a canonical transformation
and the corresponding Hamiltonian function is given by,
\begin{align}
H_{K}:=-\frac{1}{2} \omega(h,h^{(K)})= -\frac{1}{2}
\omega(\tilde{\chi},\tilde{\chi}^{(K)})
\end{align} 

For the time translation Killing vector $T$, $H_T$ (=: $E_T$), is obtained as,
\begin{eqnarray}
E_{T}:=-\frac{1}{2}\omega(\tilde{\chi},\tilde{\chi}^{(T)}) & = &
-\frac{1}{32\pi H} \mathrm {\int}_{\Sigma_{\eta}}d^{3}x
(\tilde{\chi}_{ij}{{\cal E}}_{kl}^{(T)}- {\tilde{\chi}}_{kl}^{(T)} {\cal
E}_{ij})~\delta^{ik} \delta^{jl} \\ 
& = & -\frac{1}{32\pi H} \mathrm{\int}_{\Sigma_{\eta}}d^{3}x
(\tilde{\chi}_{ij}~{\cal L}_{T}{{\cal E}}_{kl} -{\cal E}_{kl}{\cal
L}_{T}\tilde{\chi}_{ij}-3 H \tilde{\chi}_{ij}{\cal E}_{kl})~\delta^{ik}
\delta^{jl} \label{energyonsigma}
\end{eqnarray}
This integral is independent of the choice of $\Sigma_{\eta}$ and is
conveniently performed on $\mathcal{J}^+ = \Sigma_{0}$. The Killing vector $T$
also has a smooth limit to ${\cal J}^+$, $ T\Big |_{\cal J^{+}} = -H
\big(x\partial_{x}+y\partial_{y}+ z\partial_{z}\big)$ The equation
\eqref{energyonsigma} simplifies to,
\begin{align} 
E_{T} = &\frac{1}{16\pi H} \mathrm{\int}_{\mathcal{J}^+}d^{3}x ~{{\cal
E}}_{kl}~({\cal L}_{T}\tilde{\chi}_{ij}+2 H \tilde{\chi}_{ij})~\delta^{ik}
\delta^{jl}
\end{align}
Now using $({\cal L}_{T}\tilde{\chi}_{ij}+2 H \tilde{\chi}_{ij})\big|_{\cal
J^{+}}=T^{m}\partial_{m} \tilde{\chi}_{ij}$ , 
\begin{eqnarray} \label{EnergyInfty}
E_{T} & = & \frac{1}{16\pi H} \mathrm{\int}_{\mathcal{J}^+}d^{3}x ~{{\cal
E}}_{kl}~(T^{m}\partial_{m} \tilde{\chi}_{ij})~\delta^{ik} \delta^{jl}\\
& = & \frac{1}{32\pi H^{2}} \mathrm{\int}_{\mathcal{J}^+}d^{3}x
~\Big[\frac{1}{\eta}(\partial_{\eta}^{2}-\frac{1}{\eta}
\partial_{\eta})\tilde{\chi}_{kl}\Big]^{TT} ~(T^{m}\partial_{m}
\tilde{\chi}_{ij})^{TT}~\delta^{ik} \delta^{jl} \label{Energy} 
\end{eqnarray} 
In the last line we have used equation of motion and restored the $TT$ label
\cite{ABKIII}. Both ${\cal E} _{kl}$ and $T^{m}\partial_{m} \tilde{\chi}_{ij}$
have smooth limit on ${\cal J}^{+}$. 

When evaluated at the approximated solution given in (\ref{RadiativeSoln1}), the
energy flux turns out to be given by \cite{ABKIII}, 
\begin{align} \label{EnergyAtInfinity}
E_{T} = \frac{1}{8\pi }\int_{\mathcal{J}^+} d\tau~ d^{2}s \big[
\mathcal{R}_{kl} ~\mathcal{R}^{TT}_{ij} \big]\delta^{ik}\delta^{jl},
\end{align}
where, $\mathcal{R}_{ij}$ denotes the `radiation field' on ${\cal J}^+$,
expressed in terms of source moments and is given by
\begin{equation}\label{RadiationField}
\mathcal{R}_{mn}^{TT} ~ := ~ \left[\dddot{Q}_{mn} + 3H\ddot{Q}_{mn} +
2H^2\dot{Q}_{mn} + H\ddot{\bar{Q}}_{mn} + 3H^2\dot{\bar{Q}}_{mn} +
2H^3\bar{Q}_{mn}\right]^{TT}(t_{ret}) \ ,
\end{equation}
with the overdot denoting the Lie derivative $\mathcal{L}_T$.

The instantaneous {\em power} received on $\mathcal{J}^+$ at `$\tau$' is
given by,
\begin{equation}\label{PowerAtInfinity}
P(\tau) ~ := ~ \frac{1}{8\pi}\int_{S^2} d^2s \big[ \mathcal{R}^{ij}
~\mathcal{R}^{TT}_{ij} \big](-r(\tau))\ .
\end{equation}

This expression is not manifestly positive. Manifestly positive expressions
for the flux and the power are given by \cite{ABKIII}, 
\begin{eqnarray}
E_T & = & \frac{1}{2\pi}\int_{\mathcal{J}^+}d\tau~ d^2s
\left[\partial_r\mathcal{M}_{ij}^{TT}\right]\left[
\partial_r\mathcal{M}^{ij}_{TT}\right] \ , ~~~ \mbox{where} \label{TTFlux}\\
\mathcal{M}_{ij}^{TT}(\eta - r) & := & \int d^3x'T_{ij}^{TT'}(\eta - r,
\vec{x}') ~~ ; \label{TTMDefn} \\
P(\tau) & = &
\frac{1}{2\pi}\int_{S^2}d^2s\left[\partial_r\mathcal{M}_{ij}^{TT}\right]
\left[\partial_r\mathcal{M}_{TT}^{ij}\right](-r(\tau)) \label{TTPower}
\end{eqnarray}
In the definition of $\mathcal{M}_{ij}^{TT}$, the $TT'$ on the stress tensor
on the right hand side denotes transversality with respect to the $\vec{x}'$
argument. The $\mathcal{M}_{ij}^{TT}$ has no simple relation to the various
source moments and its radial derivative is distinct from the
$\mathcal{R}_{ij}^{TT}$.  For completeness, the momentum and angular momentum
fluxes are given by \cite{ABKIII},
\begin{eqnarray}
P_j & = & \frac{1}{16\pi H}\int_{\mathcal{J}^+}d^3x
\mathcal{E}^{mn}\mathcal{L}_{\xi_j}\tilde{\chi}_{mn}^{TT}  ~~ = ~~ 0 \ ;
\label{MomentumFlux}\\
J_j & = & -\frac{1}{8\pi H}\int_{\mathcal{J}^+}d^3x
\mathcal{E}^{mn}\mathcal{L}_{L_j}\tilde{\chi}^{TT}_{mn} \nonumber \\
& = & \frac{1}{4\pi}\int_{\mathcal{J}^+}d\tau d^2s\
\epsilon_{jmn}\mathcal{R}^{nl}\left[\ddot{Q}_l^{~m} + H\dot{Q}_l^{~m} +
H\dot{\bar{Q}}_l^{~m} + H^2\bar{Q}_l^{~m}\right]^{TT}
\label{AngularMomentumFlux}
\end{eqnarray}

The momentum flux is zero because the integrand is linear in $x_j$
(parity odd) and in the angular momentum flux, the second factor is
proportional to the {\em tail term}. 

\subsection{Isaacson Prescription \cite{Isaacson}}
\label{EFFECTIVEStressTensor} 
In the previous subsection we saw a definition of {\em total energy} of
radiation field of compactly supported sources in equation
(\ref{EnergyAtInfinity}). The radiated power, received at infinity, is given
in equation (\ref{PowerAtInfinity}). In this subsection we recall an
alternative framework, based on a `short wavelength expansion' \cite{MTW,
Isaacson}, for a restricted class of sources but with the benefit of a
symmetric, conserved, suitably gauge invariant {\em effective gravitational
stress tensor}. 

Conceptually, the framework is somewhat different from perturbation about a
{\em fixed}, given background solution. It is designed to construct a class of
solutions for which there exists a coordinate system in which the metric
components display two widely separated temporal/spatial scales of variation.
The slowly varying (or long wavelength $\sim L$) component is taken as the
{\em background} component and the fast (or short wavelength $\sim \lambda$)
component whose amplitude is small compared to that of the background, is
identified as the {\em ripple} component\footnote{ In the present context, $L
\sim \Lambda^{-1/2}$ while $\lambda$ could be taken as the inverse of the
characteristic frequency. The length scale $R$ denoting the extent of a
spatially compact source satisfies, $R \ll \lambda$.  }.  These statements are
manifestly coordinate dependent, but existence of a coordinate system with
sufficiently large domain admitting such an identification, itself is a
physical property.  The calculational scheme is again iterative but now allows
for both the background and the ripple components to be corrected. To make
such a separation, an {\em averaging scheme} is introduced. It splits the
Einstein equation into two separate, coupled  equations for the background and
the ripple. These equations provide a definition of the effective
gravitational stress tensor. 

For the metric of the form $g_{\mu\nu} = \bar{g}_{\mu\nu} + \epsilon
h_{\mu\nu}$, the Einstein equation to $o(\epsilon^2)$ takes the form, 
\begin{eqnarray} \label{ExpandedEqn}
R_{\mu\nu}(\bar{g} + \epsilon h) & = & \Lambda (\bar{g}_{\mu\nu} + \epsilon
h_{\mu\nu}) + 8\pi\epsilon( T_{\mu\nu} -
\frac{1}{2}g_{\mu\nu}g^{\alpha\beta}T_{\alpha\beta} ) \nonumber \\
\therefore R^{(0)}_{\mu\nu}(\bar{g}) + \epsilon R^{(1)}_{\mu\nu}(\bar{g}, h) +
\epsilon^2 R^{(2)}_{\mu\nu}(\bar{g}, h) 
& = & \Lambda (\bar{g}_{\mu\nu} + \epsilon h_{\mu\nu}) +
8\pi\left\{\frac{}{}\epsilon T_{\mu\nu} -~\frac{1}{2}(\bar{g}_{\mu\nu} +
\epsilon h_{\mu\nu}) \right. \nonumber \\
& & \left. \hspace{0.5cm}(\bar{g}^{\alpha\beta} -~\epsilon h^{\alpha\beta} +
\epsilon^2 h^{\alpha\rho}h_{\rho}^{~\beta})(\epsilon
T_{\alpha\beta})\frac{}{}\right\}
\end{eqnarray}

Introduce an averaging over an intermediate scale $\ell$, $\lambda \ll
\ell \ll L$ which satisfies the properties: (i) average of odd powers of
$h$ vanish and (ii) average of space-time divergence of tensors are
sub-leading \cite{MTW, SteinYunes}. The average of course leaves the
$L$-scale variations intact, in particular average of $g_{\mu\nu}$
equals $\bar{g}_{\mu\nu}$.  For simplicity, we will assume that the
average of matter stress tensor is zero i.e. it has only $\lambda$-scale
variations. 

Taking average of the above equation and noting that $\langle
R^{(0)}_{\mu\nu}\rangle = R^{(0)}_{\mu\nu}$ and $R^{(2)}_{\mu\nu} - \langle
R^{(2)}_{\mu\nu}\rangle \approx (R^{(2)}_{\mu\nu})_{\lambda-\mathrm{scale}}$,
the equation (\ref{ExpandedEqn}) can be separated into equation
(\ref{BackgroundEqn}) for the background and equation (\ref{RippleEqn}) for
the ripple: 
\begin{eqnarray}
8\pi T_{\mu\nu} & = & G^{(1)}_{\mu\nu} + \Lambda h_{\mu\nu} = R^{(1)}_{\mu\nu}
- \frac{1}{2}\left(\bar{g}_{\mu\nu}R^{(1)} - \bar{g}_{\mu\nu}
h^{\alpha\beta}\bar{R}_{\alpha\beta} + h_{\mu\nu}\bar{R}\right) + \Lambda
h_{\mu\nu} \label{RippleEqn} \\
8\pi t_{\mu\nu} & = & \bar{R}_{\mu\nu} - \frac{1}{2}\bar{g}_{\mu\nu}\bar{R} +
\Lambda \bar{g}_{\mu\nu} \hspace{1.0cm} with, \label{BackgroundEqn}\\
t_{\mu\nu}(\bar{g}, h) & := & ~- \frac{\epsilon^2}{8\pi}\left[ \langle
R^{(2)}_{\mu\nu}\rangle -
\frac{1}{2}\bar{g}_{\mu\nu}\bar{g}^{\alpha\beta}\langle
R^{(2)}_{\alpha\beta}\rangle \right] \label{RippleTensorDef}
\end{eqnarray}

The equation (\ref{RippleEqn}) is exactly the same linearised Einstein
equation we had before for the weak field $h_{\mu\nu}$ and every term of
it has a scale of variation $\lambda$.  However, the equation
(\ref{BackgroundEqn}) for the background is different. Although it has
terms of order $\epsilon^2$, every term has a scale of variation $L$.
If we now recognise that for $\lambda-$scale variation, $\partial h \sim
\lambda^{-1} h$ and $\epsilon' := \lambda/L$ is taken to be of the same
order as $\epsilon$, then the effective stress tensor which has a
leading term of the form {$(\partial h)^2$}, is of the order
$(\epsilon/\epsilon')^2 \sim o(1)$ and is thus included in the equation.

The effective stress tensor defined in equation (\ref{RippleTensorDef})
is manifestly symmetric and is covariantly conserved w.r.t. the
background covariant derivative, since divergence of the right hand side
of (\ref{BackgroundEqn}) vanishes identically.  For ripples over the
Minkowski background, it is gauge invariant and the energy momentum
computed using it, agrees with the quadrupole formula obtained by other
methods, thereby strengthening its interpretation as {\em gravitational}
stress tensor.  An averaging procedure constructing a tensor has been
given in \cite{Isaacson, ChoquetBruhat} {and an explicit illustrative
computation is given in the appendix}. 

At the zeroth iteration, we choose the Poincare patch of the de Sitter
space-time as the solution of (\ref{BackgroundEqn}), ignoring the effective
gravitational stress tensor.  Let us quickly verify gauge invariance of
$t_{\mu\nu}$ under $\delta_{\xi}h_{\mu\nu} = {\cal L}_{\xi}\bar{g}_{\mu\nu}$,
to leading order in $\epsilon$.  Recall that the gauge transformation involves
derivatives of $\xi_{\mu}$ and for a consistency with the background plus
ripple split, the gauge transformation should also be restricted to preserve
it.  There are two possibilities for the generator: (i) $\xi$ is comparable
with $h$ and slowly varying, and (ii) $\xi$ is order $\epsilon h$ but is
rapidly varying so that its derivative becomes order $h$. The gauge
transformation of $t_{\mu\nu}$, {after dropping space-time divergences in
the averaging, has left over} terms of the form $\Lambda \langle h\nabla \xi
\rangle$. {These vanish identically for Minkowski background making the
$t_{\mu\nu}$ gauge invariant}. For the $\xi$ of type (i), the average vanishes
since the enclosed quantity is rapidly varying and for $\xi$ of type (ii), the
averaged quantity is order $\epsilon$.  But $t_{\mu\nu}$ itself is $o(1)$ and
hence {\em gauge invariance of $t_{\mu\nu}$ is ensured to the leading order}
\cite{Isaacson}. 

Using the properties of averaged quantities, the effective gravittaional
stress tensor for  the gauges fixed solution of the ripple equation
evaluates to, 
\begin{eqnarray}\label{GaugeFreeRippleTensor}
-~8\pi t_{\mu\nu} & = & \epsilon^2\left\langle\left[ -
\frac{1}{4}\bar{\nabla}_{\mu}\tilde{h}^{\alpha\beta}
\bar{\nabla}_{\nu}\tilde{h}_{\alpha\beta} +
\frac{\Lambda}{3}\tilde{h}_{\mu}^{~\alpha}\tilde{h}_{\alpha\nu}
-\frac{\Lambda}{4}\bar{g}_{\mu\nu}\left(\tilde{h}^{\alpha\beta}
\tilde{h}_{\alpha\beta}\right) \right] \right\rangle 
\end{eqnarray}

This expression reduces to the stress tensor for the Minkowski background by
taking $\nabla_{\mu} \rightarrow \partial_{\mu}$ and dropping the last two
terms. However, for the ripple, $\partial\tilde{h} \sim \lambda^{-1}\tilde{h}
\sim \epsilon^{-1}\tilde{h}$. The connection terms in the covariant
derivatives are order $\tilde{h}$.  Hence, to the leading order in $\epsilon
\sim \lambda/L$, {\em all} terms without {\em derivatives of the ripple}, can
be dropped and we are back to the same expression for the Minkowski
background. Notice that the leading term has no $\epsilon$.

In the conformal coordinates, substituting $\tilde{h}_{\alpha\beta} =
a^2\tilde{\chi}_{\alpha\beta}$ and once again, keeping only the terms with
derivatives of the ripple, the stress tensor for the fully gauge fixed
solutions of the (\ref{RippleEqn}) becomes,
\begin{equation} \label{RippleTensorFinal}
t_{\mu\nu} ~ = ~ \frac{1}{32\pi}\langle \partial_{\mu} \tilde{\chi}_{ij}^{TT}\
\partial_{\nu} \tilde{\chi}^{ij}_{TT} \rangle \ .
\end{equation}
We will refer to this as the {\em ripple stress tensor}.  We will compute this
for the `tt' projected, approximated retarded solution, $\chi_{ij}^{tt}$. In
the subsection \ref{FROMtttoTT}, we will discuss how the computations change
when $\chi_{ij}^{tt} \to \mathcal{X}_{ij}^{TT}$.
%
\section{Conserved Quantities} \label{ENERGY}
Given {\em any} symmetric, conserved stress tensor, for every Killing vector
of the background space-time, $\xi^{\mu}$, the current $J_{\xi}^{\mu} \sim
T^{\mu}_{~\nu}\xi^{\nu}$, is covariantly conserved.  In order that for a
future directed time-like Killing vector, the corresponding energy-momentum
current is also time-like and future directed, we define $J_{\xi}^{\mu} := -
T^{\mu}_{~\nu}\xi^{\nu}$.  We adopt this definition for the time translation
Killing vector $T = -H(\eta\partial_{\eta} + x^i\partial_i)$.

The time translation Killing vector field, $T^{\nu}\partial_{\nu}$
involves only the $\eta$ and $r$ derivatives since $x^i\partial_i =
r\partial_r$ and these pass through the $\Lambda-$projector. For the space
translation along $j^{th}$ direction, we have $\partial_j$ which does
act on the $\Lambda-$projector.  In the present context where
derivatives of the ripple dominate over (ripple/r), the derivative of
the projector can be neglected and we write, $\partial_j\chi_{mn}^{tt}
\approx \hat{x}_j\partial_r\chi_{mn}^{tt}$.  For generators of rotation
however the situation is different. Once again we get two term from the
$\partial_i$, but now the $\epsilon_{jki}x^k\hat{x}^i\partial_r\chi^{mn}
~ = ~ 0!$ and we can no longer neglect the derivative of the projector.
With these understood, we write the the corresponding currents,
$J_{\xi}^{\mu} = -
\Case{a^{-2}}{32\pi}\langle\partial^{\mu}\chi^{mn}_{tt}
~\partial_{\nu}\chi_{mn}^{tt}\rangle\xi^{\nu}$. Note that the ripple
stress tensor has been defined as a covariant rank 2 tensor and hence
there is the factor of $a^{-2} = H^2\eta^2$ since the index $\mu$ has
been raised. {The currents are given by}, 
\begin{eqnarray}
a^2 J^{\eta}_{T} & = & - \frac{H}{32\pi}\left\{\eta
\langle\partial_{\eta}\chi^{mn}_{tt}~\partial_{\eta}\chi_{mn}^{tt}\rangle
+ r
\langle\partial_{\eta}\chi^{mn}_{tt}~\partial_{r}\chi_{mn}^{tt}\rangle\right\}
\\
a^2 J^{i}_{T} & = & \frac{H}{32\pi}\left\{\eta
\langle\hat{x}^i\partial_{r}\chi^{mn}_{tt}~\partial_{\eta}\chi_{mn}^{tt}\rangle
+ r
\langle\hat{x}^i\partial_{r}\chi^{mn}_{tt}~\partial_{r}\chi_{mn}^{tt}\rangle\right\}
\label{EnergyCurrent}\\
a^2 J^{\eta}_{\xi_j} & = & \frac{1}{32\pi}
\langle\partial_{\eta}\chi^{mn}_{tt}~\hat{x}_j\partial_{r}\chi_{mn}^{tt}\rangle\
~~,~~ a^2 J^{i}_{\xi_j}  =  - \frac{1}{32\pi}
\langle\hat{x}^i\partial_{r}\chi^{mn}_{tt}~\hat{x}_j\partial_{r}\chi_{mn}^{tt}\rangle\
\label{MomentumCurrent}\\
a^2 J^{\eta}_{L_j} & = & - \frac{1}{16\pi}\epsilon_{jmn}\hat{x}^m
\langle\partial_{\eta}\chi^{nl}_{tt}~\chi _{lk}\hat{x}^k\rangle ~~,~~
a^2 J^{i}_{L_j} ~ = ~ \frac{1}{16\pi}\epsilon_{jmn}\hat{x}^m
\langle\hat{x}^i\partial_r\chi^{nl}_{tt}~\chi _{lk}\hat{x}^k\rangle
\label{AngularMomentumCurrent}
\end{eqnarray} 
{The unit vectors within the angular brackets have come from the
spatial derivatives while those outside the brackets come from the
Killing vector. It is shown in the appendix (eq. \ref{AngularAverage})
that for the averaging regions far away from the source, the {\em unit
vectors can be taken across the angular brackets and we will do so in
the subsequent expressions}.}

Notice that for the energy and momentum currents (\ref{EnergyCurrent},
\ref{MomentumCurrent}), both fields have the `tt' label whereas for the
angular momentum current \eqref{AngularMomentumCurrent}, the second
factor does {\em not} have the tt label. The entire contribution to the
angular momentum current comes from the derivative of the
$\Lambda-$projector.  The contribution from the derivative of the field
vanishes since the field (without the projector) is spherically
symmetric. In all these equations we may use $\partial_r\chi_{mn} = -
\partial_{\eta}\chi_{mn} - \Case{f_{mn}}{r^2}$ from
\eqref{ChiDerivatives}. 

We note in passing that {\em if} the $\Case{f_{mn}}{r^2}$ can  be neglected
compared to $\partial_{\eta}\chi_{mn}$, then the currents corresponding to the
generators of time and space translations, both become {\em proportional} to
the vector $(1, x^i/r)$ which is a null vector.  Both energy and momentum
propagate along this direction.

Let
$\mathcal V$ denote a space-time region with a boundary
$\partial\mathcal V$. Then it follows that,
\begin{equation} \label{GeneralConservationEqn}
0 ~ = ~ \int_{\mathcal
V}d^4x\sqrt{\bar{g}}\bar{\nabla}_{\mu}J_{\xi}^{\mu} = \int_{\mathcal
V}d^4x\partial_{\mu}(\sqrt{\bar{g}}J_{\xi}^{\mu}) = \int_{\partial\cal
V}d\sigma_{\mu}J_{\xi}^{\mu} , 
\end{equation}
where $d\sigma_{\mu}$ is the oriented volume element of the boundary
$\partial\mathcal V$.

In the next subsection we evaluate the {\em energy flux},
$\mathcal{F}_{\Sigma} := \int_{\Sigma}d\sigma_{\mu}J^{\mu}_{T}$, for
various hyper-surfaces, $\Sigma$'s.  These, together with the
conservation equation (\ref{GeneralConservationEqn}) will be used to
relate power received at $\mathcal{J}^+$ to that crossing the
cosmological horizon. In the following subsection, we will present the
fluxes for momentum and angular momentum.


\subsection{Flux computations} \label{FLUXComputations}
We present flux calculations for three classes of hypersurfaces: (a)
hypersurfaces of constant physical radial distance, (b) space-like
hypersurfaces of constant $\eta$ and (c) the out-going and in-coming null
hypersurfaces. 

{\em The solution $\chi_{ij}$ in this and the next subsection stands for
$\chi_{ij}^{tt}$.}
\begin{figure}[htb]
\includegraphics[width=1.0\textwidth]{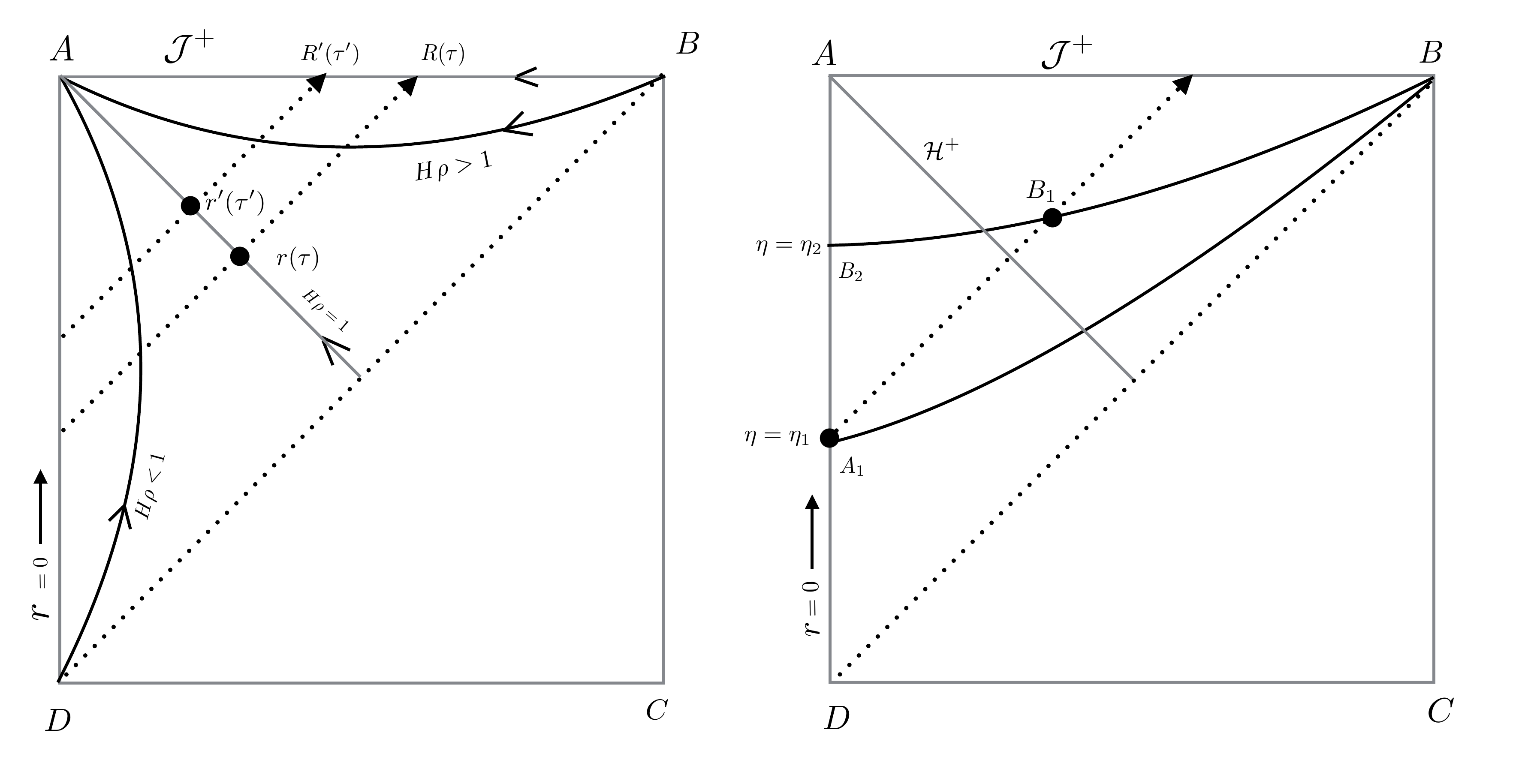}
\caption{The figure on the left shows the $\rho = \ $constant hypersurfaces
which are time-like for $H\rho < 1$, null for $H\rho = 1$ and space-like for
$H\rho > 1$. The two 45 degree out-going null hypersurfaces intersecting the
$\mathcal{H}^+$ and $\mathcal{J}^+$ in the spheres at $r(\tau), r'(\tau'),
R(\tau), R'(\tau')$, bound a space-time region. The figure on the right shows
the space-like hypersurfaces with constant value of $\eta$. The fluxes across
the out-going null hypersurfaces turn out to be zero signifying sharp
propagation of energy-momentum and angular momentum. Hence the energy flux
across the portion of the horizon bounded by the spheres at $r(\tau),
r'(\tau')$ equals the flux across the portion of the future infinity bounded
by the spheres at $R(\tau), R'(\tau')$.}\label{fluxfig}
\end{figure}
\subsubsection{Hypersurface of constant physical radial distance:}
These hypersurfaces are time-like, null and space-like according as the
physical distance being less than, equal to and greater than the physical
distance to the cosmological horizon, namely $H^{-1}$. They are spanned by the
integral curves of the Killing vector $T$.

This Killing vector is special because in the static patch, it is
time-like and its integral curves represent Killing observers. Denoting
$x^i := r\hat{x}^i,\ \hat{x}^i\hat{x}^j\delta_{ij} = 1$, in general, its
integral curves are given by $\eta(\tau) = \eta_*e^{-H\tau} ,
r(\tau) = r_*e^{-H\tau}, \hat{x}^i = \hat{x}^i_*$. Evidently,
along each curve, $\rho := r/(-H\eta) = r_*/(-H\eta_*)$ is constant.
This also represents the {\em physical radial distance}, $r_{phy} :=
|\Omega|r$. Each particular curve is labelled by $\rho$ and the two
angular coordinates $\hat{x}^i_*$. We compute the flux across the
hypersurface $\Sigma_{\rho}$, defined by $r_{phy} = \rho$. This surface
is coordinatized by the Killing parameter $\tau$ and the usual
spherical angles $\theta, \phi$ represented by the unit vectors
$\hat{x}^i$. These hypersurfaces are topologically $\Sigma_{\rho} \sim
\Delta\tau\times S^2$ and their embedding is given by,
\[
\eta(\tau, \theta, \phi) = \eta_*e^{-H\tau} ~,~ x =
r_*e^{-H\tau}sin\theta cos\phi ~,~ y = r_*e^{-H\tau}sin\theta
sin\phi ~,~ z = r_*e^{-H\tau}cos\theta \ ,
\]
with $r_* + H\rho\eta_* = 0$.

The induced metric is given by $h_{ab} = diag(H^2\rho^2 - 1, \rho^2,
\rho^2sin^2\theta)$. This has Lorentzian signature for $H\rho < 1$
(inside the static patch), is degenerate for $H\rho = 1$ (the
cosmological horizon) and Euclidean signature for $H\rho > 1$ ( beyond
the cosmological horizon). The measure factor for the non-null cases is
given by $\sqrt{|det \ h_{ab}|} = \sqrt{|1 - H^2\rho^2|}\rho^2sin\theta$
while on the cosmological horizon it is given by $\sqrt{h_{\U{ab}}} =
\rho^2sin\theta$. Here $\U{a}, \U{b}$ denote the `transverse'
coordinates $\theta, \phi$.  In the non-null case, the unit normal is
given by $n_{\mu} = \Case{\epsilon}{|H\eta|}|1 -
H^2\rho^2|^{-1/2}(H\rho, x_i/r) \leftrightarrow n^{\mu} =
\epsilon|H\eta||1 - H^2\rho^2|^{-1/2}(-H\rho, x^i/r)$. Here $\epsilon =
+1$ for {\em time-like} $\Sigma_{\rho}$ ($H\rho < 1$) and $\epsilon =
-1$ for {\em space-like} $\Sigma_{\rho}$ ($H\rho > 1$). On the
cosmological horizon, we {\em choose} the normal to be: $n_{\mu} = -
|H\eta|^{-1}(H\rho, x_i/r) \leftrightarrow n^{\mu} = -~|H\eta|( -H\rho,
x^i/r)$, so that $n^{\mu} = T^{\mu}$ is future directed. Introduce
$N^{\mu} := (- H\rho, \hat{x}^i)$, so that the normal for non-null cases
is expressed as $n^{\mu} = \epsilon|H\eta||1 - H^2\rho^2|^{-1/2}
N^{\mu}$. Note that the $n^{\mu}$ is the same for the space-like and the
null hypersurfaces, $\Sigma_{\rho \ge H^{-1}}$. For the time-like
hypersurface, the $n^{\mu}$ points in the opposite direction. However,
the induced orientation on $\Sigma_{\rho}$ is also reversed as the
hypersurface changes from being space-like to being time-like. Hence,
in {\em all cases, $H\rho > 0$}, $n^{\mu}\sqrt{h} = -
|H\eta|N^{\mu}\rho^2\ sin\theta$ and the hypersurface integral is
expressed as, 
\begin{eqnarray} 
\int_{\Sigma_{\rho}}d \Sigma_{\alpha} J_T^{\alpha} & = & -
\int_{\tau_{1}}^{\tau_{2}}d\tau\int_{S^2}d^2s\ \rho^2 
(-|H\eta(\tau)|N^{\mu})(-t_{\mu\nu}T^{\nu}) \hspace{1.0cm} \mbox{with}
\label{FluxEqn0}\\
N^{\mu}t_{\mu\nu}T^{\nu} & = &  -Hr(\tau)\left\{t_{\eta\eta} +
\hat{x}^it_{ij}\hat{x}^j -~\hat{x}^i t_{i\eta}\left((H\rho)^{-1} +
H\rho\right) \right\} \nonumber
\end{eqnarray}
The minus sign in front of the hypersurface integral is because  the
orientation defined by the Killing parameter and the angles, is {\em
negative} relative to that defined by the $r$ and the angles. The $sin
\theta$ is absorbed in $d^2s$. The minus sign in the last  parentheses is
due to the definition $J_{\mu} = - t_{\mu\nu}T^{\nu}$. In the second
line, we have also used  $-H\rho\eta = r$ valid on $\Sigma_{\rho}$. 

Substituting for the ripple stress tensor, {and taking the unit vectors
$\hat{x}$ outside of the angular bracket as mentioned before,} the expression
within the braces becomes, 
\begin{equation}\label{AveIntegrand}
\Big\{\Big\} = \frac{1}{32\pi}\left\{ \left\langle \partial_{\eta}\chi_{mn}\
\partial_{\eta}\chi^{mn} + \partial_r\chi_{mn}\
\partial_r\chi^{mn}\right\rangle - \frac{1 + H^2\rho^2}{H\rho}
\left\langle\partial_r\chi_{mn}\ \partial_{\eta}\chi^{mn}
\right\rangle\right\}
\end{equation} 
%
%
The (implicit) tt projection introduces angle dependence in the
$\chi_{ij}^{tt}$, however equation (\ref{FluxEqn0}) needs only $\eta$ and $r$
derivatives.

Eliminating $\partial_r\chi_{ij}$ using equation \eqref{ChiDerivatives},
we write,
\begin{eqnarray}
\Big\{\Big\} & = & \frac{1}{32\pi}\left[
\left\langle\partial_{\eta}\chi_{mn}\partial_{\eta}\chi^{mn}\right\rangle\frac{(1
+ H\rho)^2}{H\rho} +
\left\langle\frac{f_{mn}}{r^2}\partial_{\eta}\chi^{mn}\right\rangle\frac{(1
+ H\rho)^2}{H\rho} + \left\langle\frac{f_{mn}f^{mn}}{r^4}\right\rangle
\right] \nonumber \\
& = & \frac{1}{32\pi}\frac{(1 + H\rho)^2}{H\rho} \left[
\left\langle\partial_{\eta}\chi_{mn}\partial_{\eta}\chi^{mn} +
\frac{f_{mn}}{r^2}\partial_{\eta}\chi^{mn}\right\rangle +
\frac{H\rho}{(1 + H\rho)^2}
\left\langle\frac{f_{mn}f^{mn}}{r^4}\right\rangle \right] \nonumber \\
\therefore \int_{\Sigma_{\rho}}d \Sigma_{\alpha} J_T^{\alpha} & = &
\int_{\tau_{1}}^{\tau_{2}}d\tau\int_{S^2}d^2s\ \left[-\rho^2
H^2\eta(\tau)r(\tau)\right]\Big\{ \Big\}  \label{PhysicalFlux}
\end{eqnarray}

{The approximated solution $\chi_{ij}$, is valid for (source
dimension)/(distance to the source) $\ll 1$.  This is consistent with
the assumption that the $\lambda/r \ll 1$. Furthermore, the source being
rapidly changing, $\lambda H \ll 1$, it follows that $f_{mn}/r^2 \ll
\dot{f}_{mn}/r$.  Hence we drop $f_{mn}/r^2$ terms.  With this, $\{ \}$
takes a simple quadratic form $\Case{1}{32\pi}(1 + H\rho)^2(H\rho)^{-1}
\langle\partial_{\eta}\chi_{mn} \partial_{\eta}\chi^{mn}\rangle$. }

To compute $\partial_{\eta}$ we recall, $\eta_{ret} = \eta - r := -
H^{-1}e^{-Ht_{ret}} := - (Ha(t_{ret}))^{-1}$ and use,
\[
\partial_{\eta}f_{ij}(\eta_{ret}) =
\partial_{\eta_{ret}}f_{ij}(\eta_{ret}) =
a(\eta_{ret})\partial_{t_{ret}}f_{ij}(t_{ret}) =
a(\eta_{ret})\left(\mathcal{L}_T + 2H\right)f_{ij}(t_{ret})
\]
This leads to (overdot denoting $\mathcal{L}_T$),
\begin{eqnarray}
\partial_{\eta}\chi_{mn}(\eta_{ret}) & = &
\frac{1}{r}\partial_{\eta}f_{mn}(\eta_{ret}) +
\partial_{\eta}g_{mn}(\eta_{ret}) \nonumber \\
& = & \frac{a(t_{ret})}{r}\left(\mathcal{L}_Tf_{mn} + 2Hf_{mn}\right) +
a(t_{ret})\left(\mathcal{L}_Tg_{mn} + 2g_{mn}\right) \\
\mathcal{L}_Tf_{mn} & = & \frac{2}{a(t_{ret})}\left[\dddot{Q}_{mn} +
H\ddot{Q}_{mn} - 2H^2\dot{Q}_{mn} + H\ddot{\bar{Q}}_{mn} +
H^2\dot{\bar{Q}}_{mn} - 2H^3\bar{Q}_{mn}\right] \\
\mathcal{L}_Tg_{mn} & = & - 2H\left[\dddot{Q}_{mn} + H\ddot{Q}_{mn} +
H\ddot{\bar{Q}}_{mn} + H^2\dot{\bar{Q}}_{mn}\right] \hspace{1.0cm}
\Rightarrow\\
\partial_{\eta}\chi_{mn}^{tt}(\eta_{ret}) & = &
\frac{2}{r}\frac{\eta}{\eta - r} \mathcal{Q}_{mn}^{tt} ~~~ \mbox{with,}
\label{RttDefn}\\
\mathcal{Q}_{mn}^{tt} & := & \left[\dddot{Q}_{mn} + 3H\ddot{Q}_{mn} +
2H^2\dot{Q}_{mn} + H\ddot{\bar{Q}}_{mn} + 3H^2\dot{\bar{Q}}_{mn} +
2H^3\bar{Q}_{mn}\right]^{tt}(t_{ret}) ~~\label{RadnField}
\end{eqnarray}
Here we have also used $(1 - a(t_{ret})rH) = \Case{\eta}{\eta - r}$.
Collecting all expressions, we write the flux through a segment of
$r_{phy} = \rho$ hypersurface in a convenient form as,
{ \begin{eqnarray} 
\int_{\Sigma_{\rho}}d \Sigma_{\alpha} J_T^{\alpha} & = &
\int_{\tau_{1}}^{\tau_{2}}d\tau\int_{S^2}d^2s\left[-\rho^2
H^2\eta(\tau)r(\tau)\right]\left[\frac{(1 + H\rho)^2}{32\pi
H\rho}\right]\left\langle\left[\frac{2}{r}\frac{\eta}{\eta - r}\right]^2
\mathcal{Q}^{tt}_{ij}\mathcal{Q}_{tt}^{ij}\right\rangle \nonumber \\
& = & \int_{\tau_{1}}^{\tau_2}d\tau\int_{S^2}d^2s
\left[\frac{1}{8\pi}\right]
\left\langle\mathcal{Q}^{tt}_{ij}\mathcal{Q}_{tt}^{ij}\right\rangle(t_{ret})
\label{RhoFlux} 
\end{eqnarray}
In the appendix, we show that for large $\rho$, the expression within
the square brackets inside the angular brackets, can be taken outside.
Then, using $r = - H\rho\eta$ which is valid over the hypersurface
$\forall \rho \in \mathbb{R}^+$, we see that the explicit dependence on
$\rho$ (for large enough $\rho$) disappears from the integrand but there
is an implicit dependence on $\rho$ and $\tau$ through $t_{ret}$.} If
however, the $\tau-$integration is extended over its full range,
$(-\infty, \infty)$, then the integral {\em is} independent of $\rho$ as
well. Hence, {\em for sufficiently large $\rho$, all Killing observers
infer the same energy flux in the limit $(\tau_1, \tau_2) \to (-\infty,
\infty)$}.

The $\rho$ independence of the full flux integral in particular means
that the total flux across $\mathcal{J}^+$ equals the total flux across
the cosmological horizon, $\mathcal{H}^+$.
\begin{equation} \label{FluxHorizon}
\lim_{\rho \to \infty}\int_{\Sigma_{\rho}}d\Sigma_{\mu}J^{\mu}_T ~ = ~
\int_{\Sigma_{(H\rho = 1)}}d\Sigma_{\mu}J^{\mu}_T ~ \Leftrightarrow ~
\int_{\mathcal{J}^+}d\Sigma_{\mu}J^{\mu}_T ~ = ~
\int_{\mathcal{H}^+}d\Sigma_{\mu}J^{\mu}_T\ .
\end{equation}

\subsubsection{Flux through a constant $\eta$ slice:}
The hypersurface $\Sigma_{\eta_0}$ defined by $\eta = \eta_0$ is a
cosmological slice $\sim \mathbb{R}^3$. It is space-like, with a normal
$n_{\mu} = -|H\eta_0|^{-1}(1, \vec{0}) \leftrightarrow n^{\mu} =
|H\eta_0|(1, \vec{0})$ which is future directed. We choose a finite
portion of it with $r \in [r_1, r_2]$. The hypersurface is topologically
$\Delta r\times s^2$. Choosing the $(r, \theta, \phi)$ coordinates on
the hypersurface, the embedding is given by 
\[
\eta(r, \theta, \phi) = \eta_0~,~ x = r\ sin\theta\ cos\phi ~,~ y = r\
sin\theta\ sin\phi ~,~ z = r\ cos\theta \ .
\]
The induced metric is given by $h_{ab} = (H\eta_0)^{-2}diag(1, r^2,
r^2sin^2\theta)$ giving $\sqrt{|det\;h_{ab}|} = |H\eta_0|^{-3}r^2
sin\theta$. Denoting $N^{\mu} := (1, \vec{0})$, the hypersurface
integral is given by,
\begin{eqnarray}
\int_{\Sigma_{\eta_0}}d\Sigma_{\mu}J^{\mu}_T & = &
\int_{r_1}^{r_2}dr\int_{S^2}d^2s\ r^2
a^2(\eta_0)\left(-N^{\mu}t_{\mu\nu}T^{\nu}\right) \hspace{1.0cm}
\mbox{with}\\
N^{\mu}t_{\mu\nu}T^{\nu} & = & (-H)\left(t_{\eta\eta}\eta + t_{\eta
i}x^i\right) \nonumber \\
& = &
\frac{-H}{32\pi}\left(\eta\langle\partial_{\eta}\chi_{mn}\partial_{\eta}\chi^{mn}\rangle
+ x^i\langle\partial_{\eta}\chi_{mn}\ \partial_i\chi^{mn}\rangle \right)
\\
& = & \frac{-H}{32\pi}\left((\eta -
r)\langle\partial_{\eta}\chi_{mn}\partial_{\eta}\chi^{mn}\rangle -
\left\langle\frac{f_{mn}}{r}\partial_{\eta}\chi^{mn}\right\rangle\right)
\\
\therefore \int_{\Sigma_{\eta_0}}d\Sigma_{\mu}J^{\mu}_T & = &
-\frac{1}{32\pi H^2\eta_0^2}\int_{r_1}^{r_2}dr\int_{S^2}d^2s\ r^2
\left\{\frac{1}{a(\eta_{ret})}\langle\partial_{\eta}\chi_{mn}
\partial_{\eta}\chi^{mn}\rangle\right\} 
\\
& \approx & \int_{r_1}^{r_2}\frac{dr}{H(r - \eta_0)}\int_{S^2}d^2s\
\left[\frac{-1}{8\pi}\right]\langle\mathcal{Q}_{mn}^{tt}\mathcal{Q}^{mn}_{tt}\rangle
\label{EtaFlux}
\end{eqnarray}
By the same reasoning as before, we have dropped the $\Case{f_{mn}}{r}$
and also used equation (\ref{TEtaEta}).  In the limit $\eta_0 \to 0$
with $(r_1, r_2) \to (0, \infty)$, the hypersurface becomes
$\mathcal{J}^+$ and the integration measure becomes $\Case{dr}{Hr}$. The
limit $\eta \to 0$ is thus finite.

As noted earlier, the hypersurface integral when expressed in terms of
the Killing parameter, has a minus sign due to the reversal of the
induced orientation. The measures (positive) themselves are related as
$\Case{dr}{Hr} = d\tau$, leading to $\int_0^{\infty}dr/Hr = -
\int_{-\infty}^{\infty}d\tau$ and we get, 
\begin{equation}
\lim_{\eta_0 \to 0}\int_{\Sigma_{\eta_0}}d\Sigma_{\mu}J^{\mu}_T =
\lim_{\rho \to \infty}\int_{\Sigma_{\rho}}d\Sigma_{\mu}J^{\mu}_T \ .
\end{equation}

\subsubsection{Flux through null hypersurfaces:}
There are two families of {\em future directed} null hypersurfaces given
by $\eta + \epsilon r + \sigma = 0$, see figure (\ref{fluxfig}). For
$\epsilon = +1$, these 45 degree lines in the Penrose diagram are
parallel to the cosmological horizon while for $\epsilon = -1$, the
lines are parallel to the null boundary of the Poincare patch. We refer
to these as the {\em in-coming} ($\epsilon = 1$) and {\em out-going}
($\epsilon = -1$) null hypersurfaces.  The parameter $\sigma$ labels
members of these families.

The null normals of these families are of the form $n_{\mu} = \gamma (1,
\epsilon\hat{x}_i) \leftrightarrow n^{\mu} = (H\eta)^2\gamma(-1,
\epsilon\hat{x}^i)$, where $\gamma$ is to be chosen suitably and should
be {\em negative} for future directed hypersurfaces. Choosing
coordinates $(\lambda, \theta, \phi)$ on a null hypersurface, its
embedding may be taken as $\eta(\lambda), r(\lambda)$ with identity
mapping of the angles. Here $\lambda$ is an affine parameter of the null
geodesics generating the null hypersurfaces. The induced metric is
obtained as $h_{ab} = (H\eta)^{-2}diag(0, r^2, r^2 sin^2\theta)$. Note
that the orientation of the hypersurfaces, relative to that defined by
$(r, \theta, \phi)$ is the same for the out-going hypersurfaces and
opposite for the in-coming hypersurfaces. The hypersurface integral is
then given by ($N^{\mu} := (-1, \epsilon\hat{x}^i$)),
\begin{eqnarray}
\int_{\Sigma_{(\epsilon, \sigma)}}d\Sigma_{\mu}J^{\mu}_T & = & -\epsilon
\int^{\lambda_2}_{\lambda_1}d\lambda\int_S^2d^2s
\left[\frac{r^2}{H^2\eta^2}\right] (H^2\eta^2)\gamma
(-N^{\mu}t_{\mu\nu}T^{\nu}) ~~, \\
N^{\mu}t_{\mu\nu}T^{\nu} & = & - H\left( -t_{\eta\eta}\eta - t_{\eta
j}r\hat{x}^j + \epsilon\hat{x}^it_{i\eta}\eta + \epsilon r
\hat{x}^it_{ij}\hat{x}^j \right) \nonumber \\
& = & -\frac{H}{32\pi}\left(
-\eta\langle\partial_{\eta}\chi_{mn}\partial_{\eta}\chi^{mn}\rangle +
(\epsilon\eta -
r)\langle\partial_{\eta}\chi_{mn}\partial_r\chi^{mn}\rangle + \epsilon r
\langle\partial_r\chi_{mn}\partial_r\chi^{mn}\rangle\right) \nonumber \\
& = & -\frac{H}{32\pi}(1 + \epsilon)(r -
\eta)\langle\partial_{\eta}\chi^{tt}_{mn}\partial_{\eta}\chi_{tt}^{mn}\rangle
+ o(r^{-2}) \\
& = & -\frac{H}{32\pi}(1 + \epsilon)(r -
\eta)\left[\frac{4}{r^2}\frac{\eta^2}{(\eta -
r)^2}\right]\langle\mathcal{Q}^{tt}_{mn}\mathcal{Q}_{tt}^{mn}\rangle +
o(r^{-2}) \\
\therefore \int_{\Sigma_{(\epsilon, \sigma)}}d\Sigma_{\mu}J^{\mu}_T & =
& \epsilon\int^{\lambda_2}_{\lambda_1}d\lambda\int_{S^2}d^2s\left[\gamma
H\right]\left[ \frac{(1 + \epsilon)}{8\pi}\frac{\eta^2}{\eta - r}\right]
\langle\mathcal{Q}^{tt}_{mn}\mathcal{Q}_{tt}^{mn}\rangle
\label{NullFlux}
\end{eqnarray}
As before, we have dropped the $f_{mn}/r^2$ terms from
$\partial_r\chi_{mn}$ and used the equation (\ref{TEtaEta}).

It is immediately clear that the flux through the out-going null
hypersurfaces ($\rho$ or $r$ increase along these) vanishes. In the
$\epsilon = 1$ family, only the cosmological horizon is of interest. For
this we have $(\eta = -r)$ and we {\em choose} the factor $\gamma =
-(Hr)^{-1}$ so the null normal matches with the Killing vector ($\gamma$
is negative as desired for future orientation) and the affine parameter
$\lambda$ matches with the Killing parameter $\tau$. With this choice,
the flux in eqn. (\ref{NullFlux}) matches with that given in eq.
(\ref{RhoFlux}) for $H\rho = 1$.  Thus, once again, the full flux
through cosmological horizon is exactly same as that of $r_{physical}=
const$ hypersurfaces. 

{\em Remarks:} All three calculations consistently have the same
$[+1/8\pi]$ factor, with integrals oriented along the stationary Killing
vector.

It is surprising at first that the flux through $\eta-r=constant$
hypersurfaces is zero, which indicates sharp propagation of the energy,
even though the retarded solution has a tail contribution. This can be
seen more directly as follows.  Let us recast eqn.  \eqref{PhysicalFlux}
as
\begin{align}
\int_{\Sigma_{\rho}}d \Sigma_{\alpha} J_T^{\alpha}  =
\dfrac{1}{32\pi}\int_{\tau_{1}}^{\tau_{2}}d\tau\int_{S^2}d^2s ~
\bigg\langle
~r^{2}(\eta-r)^{2}~\left(\frac{1}{\eta}\partial_{\eta}\chi_{ij}\right)~
\left(\frac{1}{\eta}\partial_{\eta}\chi_{ij} \right)\bigg\rangle
\end{align}
where we have neglected the $1/r^{2}$ terms and have used $r=-H\rho
\eta$. 

Now in taking the $\eta-$derivative, contribution of the tail term in
\eqref{RadiativeSoln} cancels out, leaving only the contribution from
the sharp term:
\begin{align}
\frac{1}{\eta}\partial_{\eta}\chi_{ij}=\frac{4}{r~\eta_{ret}}
\partial_{\eta}\int d^3x' T_{ij}~(\eta-r, x') \ .
\end{align}

\subsection{Momentum and angular momentum fluxes} \label{MOMENTUMFluxes}
For the same three classes of hypersurfaces, we present the momentum and
angular momentum fluxes. We already have the measures for these
hypersurfaces as well as the currents given in (\ref{MomentumCurrent},
\ref{AngularMomentumCurrent}). The full fluxes, only to the leading
order in $r^{-1}$, are given by,
\begin{eqnarray}
\Sigma_{\rho} & : & - \int_{-\infty}^{\infty}d\tau\int_{S^2}d^2s
\rho^2\frac{1}{H\eta}(H\rho, \hat{x}_i)J^{\mu} \\
\Sigma_{\eta_0} & : & \int_{0}^{\infty}dr\int_{S^2}d^2s \left[\frac{r^2}
{|H\eta_0|^3}\right]\left[\frac{1}{H\eta_0}\right](1, \vec{0})J^{\mu} \\
\Sigma_{(\epsilon, \sigma)} & : & -\epsilon
\int_{\lambda_1}^{\lambda_2}d\lambda\int_{S^2}d^2s
\left[\frac{r^2\gamma}{|H\eta_0|^{2}}\right](1,
\epsilon\hat{x}_i)J^{\mu} 
\end{eqnarray}

{\em Momentum fluxes:} The momentum current is given by,
\begin{equation}
J^{\mu}_{\xi_j} ~ = ~ - \frac{a^{-2}}{32\pi}\left[\frac{2}{r}\frac{\eta}{\eta
-r}\right]^2\langle\mathcal{Q}^{mn}_{tt}~
\mathcal{Q}^{tt}_{mn}\rangle(\hat{x}_j)(1, -\hat{x}^i)
\end{equation}
Dotting with the $n_{\mu}$ produces a rotational scalar and the average
is a rotational scalar too. Then the angular integration with
$\hat{x}_j$ vanishes, in all three cases. Hence, the momentum flux is
zero across the three classes of hypersurfaces.

{\em Angular Momentum fluxes:} Replacing $\partial_r\chi^{mn}_{tt}
\approx - \partial_{\eta}\chi^{mn}_{tt}$, we can write the angular
momentum current as,
\[
J^{\mu}_{L_j} ~ = ~ -\frac{a^{-2}}{16\pi}\left[\frac{2}{r} \frac{\eta}{\eta -
r}\right]
\left[\epsilon_{jmn}\hat{x}^m\hat{x}^k\langle\mathcal{Q}^{nl}_{tt}~\chi_{kl}\rangle\right]
(1, \hat{x}^i)
\]

The fluxes then take the form,
\begin{eqnarray}
\Sigma_{\rho} & : & - \frac{\rho}{8\pi}\int_{-\infty}^{\infty}d\tau
\int_{S^2}d^2s \left[\epsilon_{jmn}\hat{x}^m\hat{x}^k
\langle\mathcal{Q}^{nl}_{tt}~\chi_{kl}\rangle\right] \\
\Sigma_{\eta_0} & : & \frac{1}{8\pi H^2|\eta_0|}
\int_0^{\infty}dr\frac{r}{r -\eta_0} \int_{S^2}d^2s
\left[\epsilon_{jmn}\hat{x}^m\hat{x}^k
\langle\mathcal{Q}^{nl}_{tt}~\chi_{kl}\rangle\right] \\
\Sigma_{(\epsilon, \sigma)} & : & -\epsilon\frac{1 + \epsilon}{8\pi}
\int_{\lambda_{1}}^{\lambda^2}d\lambda (-\gamma) \frac{r\ \eta}{ \eta
-~r}\int_{S^2}d^2s \left[\epsilon_{jmn}\hat{x}^m\hat{x}^k
\langle\mathcal{Q}^{nl}_{tt}~\chi_{kl}\rangle\right] 
\end{eqnarray}

{Consider the average. The function enclosed in averaging is
product of the $\Lambda-$projector containing angular dependence and a
function having dependence on $(\eta, r)$. The averaging can then be
split into averaging over {a cell $\Delta\omega$ in the angular
coordinates around the direction $\hat{r}$} and averaging over a cell in
the $(\eta, r)$ plane, see equation (\ref{AngularAverage}). Thus, we
write, 
\begin{eqnarray}
\langle\mathcal{Q}^{nl}_{tt}~\chi_{kl}\rangle(\eta, r, \hat{r}) & = &
\left[\frac{1}{\Delta\omega}\int_{\Delta\omega}d^2s'
\Lambda^{nl}_{~~~rs}(\hat{r}')\right]
\left[\langle\mathcal{Q}^{rs}\chi_{kl}\rangle(\eta, r)\right] \\
& = & \Lambda^{nl}_{~~~rs}(\hat{r})
\langle\mathcal{Q}^{rs}\chi_{kl}\rangle(\eta, r) 
\end{eqnarray}
The angular integration over the sphere can be done explicitly:
\begin{equation}
\int_{S^2}d^2s\epsilon_{jmn}\hat{x}^m\hat{x}^k
\Lambda^{nl}_{~~~rs}(\hat{r}) \langle\mathcal{Q}^{rs}\chi_{kl}(\eta,
r)\rangle ~=~ \frac{8\pi}{15}\epsilon_{jmn}
\langle\mathcal{Q}^{nl}\chi_{l}^{~m}\rangle(\eta, r)  .
\end{equation}
This is to be integrated over the Killing parameter $\tau$ or $r$ or
$\lambda$ for the three classes of hypersurfaces. The average is now
over an $(\eta, r)$ cell.  }

This integration in the flux expressions above, can be expressed in
terms of the Killing parameter $\tau$ and then they all take the same
form {\em provided} for $\Sigma_{\eta_0}$ we consider the $\eta_0
\approx 0 \to \mathcal{J}^+$ and for the null hypersurface we choose the
cosmological horizon, $\mathcal{H}^+$ ($\epsilon = +1, \eta = -r)$:
\begin{equation}
(\mbox{Flux of the of (Angular Momentum)$_j$}) ~ = ~ -
\frac{1}{15}\int_{-\infty}^{\infty}d\tau~ a(\eta(\tau))\
r(\tau)\epsilon_{jmn}\langle\mathcal{Q}^{nl}\chi_l^{~m}\rangle
\end{equation}
The radiation field $\mathcal{Q}^{nl}$ is given in equation
(\ref{RadiationField}) but without the $tt$ label and,
\begin{eqnarray}
\chi_{lm} & = & \frac{2}{r a(\eta)}\left[\ddot{Q}_{lm} + 2H\dot{Q}_{lm}
+ H\dot{\bar{Q}}_{lm} + 2H^2\bar{Q}_{lm}\right](\eta_{ret}) \nonumber \\
& & + 2H^2\left[\dot{Q}_{lm} + H\bar{Q}_{lm}\right](\eta_{ret}) -
2H^2\left[\dot{Q}_{lm} + H\bar{Q}_{lm}\right](-\infty) \ .
\end{eqnarray}

This flux does {\em not} have a finite limit to $\mathcal{J}^+$ due to
the {\em tail term} in $\chi_l^{~m}$ and does {\em not} match with the
flux given by \cite{ABKIII}. It is finite along the $\mathcal{H}^+$
though. It does {\em not} match with the correct angular momentum flux
in the flat space limit as well and it is well known \cite{MTW,
PoissonWill} that the Isaacson effective stress tensor does not suffice
to capture the flux of angular momentum.  The sharp propagation property
still holds in the sense that the flux across out-going null
hypersurface is zero.

\subsection{Extending from `tt' to `TT'} \label{FROMtttoTT}
{We have used the algebraic `tt' projection on the approximated,
retarded solution. How would the results change if we were to use the
`TT' decomposition of the exact solution prior to the
$|\vec{x}'|/|\vec{x}| \ll 1$ approximation?  For this we note a few
points.

It is easy to see that the TT part of the retarded solution is given by
\cite{ABKIII},
\begin{eqnarray}
\mathcal{X}^{TT}_{ij}(\eta, x) & = & 4 \int d^3x' \frac{\eta}{|x - x'|(\eta -
|x - x'|)} \left.  T^{TT'}_{ij}(\eta', x') \right|_{\eta' = \eta - |x - x'|}
\nonumber \\
& & \hspace{0.5cm} 
+ ~ 4 \int d^3x' \int_{- \infty}^{\eta - |x
-x'|}d\eta'\frac{T^{TT'}_{ij}(\eta', x')}{\eta'^2} \label{RadiativeSolnTT} 
\end{eqnarray}
where the $TT'$ refers to the {\em second} argument of the stress tensor. This
follows by checking that the divergence, $\partial^i_x$, of the right hand
side converts into the divergence, $\partial^i_{x'}$ on the second argument of
the stress tensor. For this relation, it is important to have the exact $|x -
x'|$ dependence and that the source has compact support. The TT part of
the approximated solution cannot be similarly expressed in terms of TT part of
the source stress tensor. 

We can now consider the solution (\ref{RadiativeSolnTT}) for $|x| \gg
|x'|$, and replace $|x - x'| \approx r$ which simplifies the source
integral. We denote this approximated expression as $\chi^{TT}_{ij}$.
This satisfies the transversality condition to $o(r^{-1})$
only\footnote{ {Extracting the TT part and making the approximation for
$|\vec{x}| \gg |\vec{x'}|$, do not commute i.e.
$[(\mathcal{X}_{ij})_{approx}]^{TT} \neq
([\mathcal{X}_{ij}]^{TT})_{approx}$. This is so because the $\partial^j$
of the l.h.s. is always zero by definition while that of the r.h.s. is
non-zero in general. We are using the r.h.s.}\label{TTApproxCommute} }.
Furthermore, since the transverse, traceless part of the stress tensor
{\em drops out of its conservation equation}, we cannot directly express
$\int_{source}T^{TT'}_{ij}$ in terms of correspondingly defined moments.
Nevertheless, we do get,
\begin{eqnarray}
\partial_{\eta}\chi^{TT}_{ij}(\eta, x) & = & 4 \frac{\eta}{r(\eta
-~r)}\partial_{\eta}\mathcal{M}_{ij}^{TT}~~,~~
\mathcal{M}_{ij}^{TT}(\eta -~r) := \int d^3x'T_{ij}^{TT'}(\eta - r, x')
\label{RTTDefn} \\
\partial_m\chi_{ij}^{TT}(\eta, x) & = &
4\frac{\hat{x}_m}{r}\left(\frac{\eta}{\eta -
r}\partial_r\mathcal{M}_{ij}^{TT}
- \frac{\mathcal{M}_{ij}^{TT}}{r}\right) ~ = ~ - \hat{x}_m
  \partial_{\eta}\chi_{ij}^{TT} - 4
  \frac{\hat{x}_m}{r^2}\mathcal{M}_{ij}^{TT} \\
\therefore \partial_r\chi_{ij}^{TT} & = & -
\partial_{\eta}\chi_{ij}^{TT} - 4 \frac{\mathcal{M}_{ij}^{TT}}{r^2}
\label{ChiTTDerivatives}
\end{eqnarray}
The equation (\ref{ChiTTDerivatives}) has the same form as
eq.(\ref{ChiDerivatives}).  The equation (\ref{RTTDefn}) has the same
form as eq.(\ref{RttDefn}) which introduced the radiation field
$\mathcal{Q}_{ij}^{tt}$. We can thus introduce a new `radiation field',
$\mathcal{Q}_{ij}^{TT} := 2\partial_{\eta}\mathcal{M}_{ij}^{TT}$.
With this, the form of the expressions for fluxes will remain the same
with $\mathcal{Q}_{ij}^{tt} \to \mathcal{Q}_{ij}^{TT}$. Note that
unlike $\mathcal{Q}_{ij}^{tt}$, the $\mathcal{Q}_{ij}^{TT}$ does
{\em not} have a simple relation to the source moments defined earlier.
Nevertheless, it shares the important property with
$\mathcal{Q}_{ij}^{tt}$, namely, it too is a function of $\eta - r$
alone. This enables the space-time averaging to be reduced to averaging
over $\rho = $constant hypersurfaces, as shown in eqn.
(\ref{TTEtaEta}).
\begin{eqnarray} \label{TTAveraged}
\langle\partial_{\eta}\chi_{mn}^{TT}\partial_{\eta}\chi_{TT}^{mn}\rangle
(t, r, \hat{r}) 
& = & 4\frac{a^2(\bar{t}_0)}{\rho_0^2} \langle
\mathcal{Q}_{ij}^{TT}\ \mathcal{Q}^{ij}_{TT}\rangle
(\bar{t}_0, \hat{r})
\end{eqnarray}

}

In the next section we restrict to the energy fluxes and see two
applications of the conservation equation and the sharp propagation
property. 

\section{Implications of conservation equation and sharp propagation}
\label{COSMOLOGICALHorizon}

In the previous subsection, we assembled fluxes through various
hypersurfaces, all having the topology $\Delta\times S^2$. We considered
$\Delta$ to be a finite interval and also the cases with $\Delta =
\mathbb{R}$. The relevant hypersurfaces have $\rho =~$constant. In all
cases, the energy flux integral had the form,
\begin{equation} \label{GeneralFlux}
\mathcal{F}(a, b) :=
\int_a^bd\tau\int_{S^2}d^2s\left[\frac{1}{8\pi}\right]
\langle\mathcal{Q}_{mn}^{tt}\mathcal{Q}^{mn}_{tt}\rangle ~ =: ~
\int_a^bd\tau \langle F\rangle (\tau) \ .
\end{equation} 
As shown in the appendix, equations (\ref{TEtaEta}), the angular
brackets denote averaging over $\tau -$intervals and a trivial averaging
over the angular intervals. Since the angular average is trivial, we
have taken the angular integration across the averaging and denoted the
integration over the sphere by $\langle F\rangle(\tau)$. Using the mean
value theorem, we write,
\begin{equation}
\int_a^bd\tau \langle F\rangle(\tau) = \langle F\rangle(c)(b -a) =
\int_{c - \delta}^{c + \delta}d\tau \ F~ \frac{(b - a)}{2\delta} ~~~,~~~
c \in (a, b) \ .
\end{equation}

Let us choose $(a, b)$ to be an averaging interval i.e. $(b - a) =
2\delta$.  Recall that the averaged quantities are slowly varying i.e.
$\langle F\rangle$ is varying only over the scale $L \gg 2\delta$ and
thus essentially constant over the averaging interval. Therefore, we can
{\em choose} $c = (a + b)/2$ possibly making a small error. But then the
right hand side of the last equality in the above equation becomes
$\int_a^b d\tau F(\tau)$. In effect, for integral over an averaging
interval, we can drop the angular brackets in equation
(\ref{GeneralFlux}).

For $a \ll 0, b \gg 0$, the $\tau$ integral can be replaced by a sum
with each sub-interval, $[a_k, b_k]$ being an averaging interval.  Using
the above argument, we can write,
\begin{equation} \label{CoarseGrained}
\mathcal{F}(a,b) \approx \sum_k
\int_{a_k}^{b_k}d\tau\int_{S^2}d^2s\left[\frac{1}{8\pi}\right]
\mathcal{Q}^{tt}_{mn}\mathcal{Q}^{mn}_{tt}
\end{equation}
However, the averaging $\tau-$intervals cannot be made arbitrarily finer
and the Riemann sum cannot be taken to the integral. Hence, flux
integral over an averaged integrand matches with the flux integral over
an {\em un-}averaged integrand only at a {\em coarse grained level}.
The same arguments also hold for $\mathcal{Q}_{mn}^{tt} \to
\mathcal{Q}_{mn}^{TT}$ and {\em then} the fluxes defined using the
averaged stress tensor match with the expressions (\ref{TTFlux}) at a
{\em coarse grained level}.  

By judicious choices of hypersurfaces comprising the boundary
$\partial\mathcal{V}$ of a space-time region $\mathcal{V}$, we can
relate different fluxes using the conservation equation
\eqref{GeneralConservationEqn}. The sharp propagation of energy comes in
very useful. We note two of its implications.

(1) The flux across two hypersurfaces $\Sigma_{\eta_1}$ and
$\Sigma_{\eta_2}$ {\em cannot} be equal, see the right side figure of
(\ref{fluxfig}). 

Let $\eta_2 > \eta_1$. Let $\Sigma_{\eta_1}$ meet the $r = 0$ line at
$A_1$. Let the out-going null hypersurface through $A_1$ intersect the
$\Sigma_{\eta_2}$ in a $S^2$ at $B_1$ with the radial coordinate being
$r_1$. The three hypersurfaces $\Sigma_{\eta_1}$, the out-going null
hypersurface and the hypersurface $\Sigma_{\eta_2}$ bounded by the
sphere at $B_1$ enclose a space-time region, $A_1 B B_1$. By the
conservation equation \eqref{GeneralConservationEqn}, the sum of the
fluxes through these bounding hypersurfaces must vanish. But the flux
through the out-going null hypersurface vanishes as shown before. Hence
the fluxes through $\Sigma_{\eta_1}$ and the partial hypersurface
$\Sigma_{\eta_2}$ between $B_1$ and $B$, must be equal.  However, this
leaves the contribution of the flux through the `remaining' portion of
the $\Sigma_{\eta_2}$ hypersurface between $B_2$ and $B_1$. Hence the
result.  Alternatively, one can also see this explicitly by writing the
full flux through the two hypersurfaces using the expression given in
equation \eqref{EtaFlux} and matching the integrands along the out-going
null hypersurface.  Evidently, the full flux through $\Sigma_{\eta \neq
0}$ is also not equal to that through $\mathcal{J}^+$. Physically this
is understandable since the hypersurface at a later value of $\eta$
receives energy emitted {\em after} the earlier value of $\eta$. The
null infinity of course records {\em all} the energy emitted by the
source and so does the cosmological horizon. We also conclude that the
total flux at $\mathcal{J}^+$ computed by Ashtekar et al, as given in
eq.(\ref{TTFlux}), matches (at coarse grained level) with that given in
equation \eqref{EtaFlux} (with $\mathcal{Q} \to \mathcal{R}$) {\em only}
for $\eta = 0$.  Note that unlike the spatial slices $\Sigma_{\eta}$,
all hypersurfaces $\Sigma_{\rho > 0}$ intercept all the emitted energy.

(2) The sharp propagation of energy can also be used to infer the {\em
instantaneous emitted power}. Consider two out-going null hypersurfaces
intersecting the cosmological horizon in spheres with radii  $r(\tau)$
and $r'(\tau')$. The same hypersurfaces intersect the null infinity at
corresponding spheres at $R(\tau)$ and $R'(\tau')$, see the left side
figure in (\ref{fluxfig}). For $\tau' > \tau$, we have $r'(\tau') <
r(\tau)$ and $R'(\tau') < R(\tau)$.  By the conservation equation and
sharp propagation, the flux integral over the portion bounded by the
spheres $R, R'$ on $\mathcal{J}^+$ and the flux integral over the
portion bounded by the spheres $r(\tau), r'(\tau')$ on the
$\mathcal{H}^+$, are equal. Taking $\tau' = \tau + \delta\tau$, the
integral becomes $\delta\tau \times $ the integral over the sphere at
$r(\tau)$. The {\em emitted power} is then defined by dividing the flux
integral by $\delta\tau$ and taking the limit.  Thus we get the
instantaneous power as:
\begin{equation}\label{PowerAtHorizon}
\mathcal{P}(\tau) ~ := ~ \lim_{\delta\tau \to 0} \frac{\mathcal{F}(\tau
+ \delta\tau, \tau)}{\delta\tau} ~ = ~ \frac{1}{8\pi}\int_{S^2}d^2s
\langle\mathcal{Q}_{ij}^{TT}\mathcal{Q}^{ij}_{TT}\rangle \ . 
\end{equation}
This is manifestly positive.

{This is very similar to the definition given by Ashtekar et al
\cite{ABKIII} in the form of equation (\ref{TTPower})
{\em except that} the integrand is an average over $\tau$ and angular
windows.  The power is usually averaged over a few periods.  If this is
done to the power expression in \cite{ABKIII}, it will match with the
above expression, again at a coarse grained level. 

The upshot is that the quadrupole power defined above is gauge invariant
and can be computed at the cosmological horizon.
} 

\section{Discussion and Summary}\label{FINALSection} 
We have dealt with two aspects namely the role of the cosmological
horizon and the use of ripple stress tensor in the limited context of
rapidly changing, distant sources. 

{A question regarding the validity of the `short wavelength
approximation' near $\mathcal{J}^+$ arises due to the understanding that
the physical wavelength will diverge near the future null infinity
thanks to the scale factor $a(t)$. Let us recall that background plus
ripple decomposition is premised over the expectation:
$\partial_{\alpha}\bar{g}_{\mu\nu} \sim \bar{g}_{\mu\nu}/L$ and
$\partial_{\alpha}h_{\mu\nu} \sim h_{\mu\nu}/\lambda$. In the
cosmological chart, the non-zero coordinate derivatives of the
background are: $\partial_{t}\bar{g}_{ij} = 2H\bar{g}_{ij} \sim
\bar{g}_{ij}/L$. For the retarded solution we have,
\begin{eqnarray}
\frac{\partial_th_{ij}}{h_{ij}} & = & \partial_t [\ell
n(a^2(t)\chi_{ij})] = 2H +  \partial_t\eta\partial_{\eta}\ell
n(\chi_{ij})  = 2H + \frac{\partial_{\eta}\ell n(\chi_{ij})}{a(t)}  \sim
\frac{1}{L} + \frac{1}{a(t)\lambda}  \ , \\
\frac{\partial_k h_{ij}}{h_{ij}} & = & \partial_k\ell n(\chi_{ij}) =
\hat{r}_k\partial_r\ell n(\chi_{ij}) \approx -
\hat{r}_k\partial_{\eta}\ell n(\chi_{ij}) \sim \frac{\hat{r}_k}{\lambda}
\end{eqnarray}

The first equation shows that the $t-$derivative of the perturbation
does {\em not} satisfy the premise, near $\mathcal{J}^+$ thanks to the
presence of the scale factor.  The second equation however does {\em
not} have the scale factor and the ripple indeed has short scale of
spatial variation. Interestingly, in the calculation of the fluxes,
spatial components of the ripple stress tensor (and hence the spatial
derivatives of the perturbation) do contribute since all Killing vectors
are space-like near $\mathcal{J}^+$ and the `short wavelength
approximation' can justifiably be used. 
}

As noted in the introduction, the cosmological horizon is unambiguously
defined for a spatially compact source. This follows because worldlines
with finite physical separation at every $\eta$ must converge to $i^+$,
the point $A$ of figure (\ref{DeSitterPenrose}). If $\Delta$ denotes the
physical radial distance  corresponding to the radial coordinate
difference $\delta$, then $\Delta^2 = \Case{\delta^2}{H^2\eta^2}$. To
maintain $\Delta^2$ to be finite as $\eta \to 0_-$, we must have
$\delta^2 \sim \alpha^2 \eta^2 + 0(|\eta|^3)$ near $i^+$. This
identifies $\delta$ with $-\alpha\eta$ or $\alpha = H\rho$. Thus, the
worldlines approach $i^+$ along the $\rho = $constant hypersurfaces. The
cosmological horizon is then the past lightcone of $i^+$.  The same
argument also shows that any observer, who remains at finite physical
distance away from the source must remain confined within the
cosmological horizon. Furthermore, neither any such observer, nor the
source has any access to energy/momentum which has crossed the horizon.
Hence cosmological horizon does share physically relevant properties
with the future infinity.  Incidentally, any future directed causal
curve reaching $\mathcal{J}^+$ also registers on $\mathcal{H}^+$. 

Further support for the role of cosmological horizon as future null
infinity comes from the computations of the energy momentum fluxes.  For
these, we employed the effective ripple stress tensor and showed that
the fluxes defined at $\mathcal{J}^+$ match with those defined at
$\mathcal{H}^+$.  Furthermore, these fluxes also matched (at a coarse
grained level) with the energy momentum fluxes defined by the more
geometric methods of the covariant phase space framework. This provides
a further support to the utility of the ripple stress tensor.  The
quadrupole power too matches likewise. The ripple stress tensor,
although limited to short wavelength regimes (which covers most common
sources), provides a convenient picture of energy momentum flows much
like the flows for matter.  There is a shortcoming of the ripple stress
tensor - it does not capture the angular momentum flux correctly. A
clearer understanding of this failure is lacking at present.

It should be noted that definition of fluxes is not necessarily {\em
unique}. Apart from a definition being well defined, its `correctness'
should be tested in conjunction with the definition of the Bondi-type
quantities having a loss formula relating to flux. Recent work within a
Bondi-type framework may be seen in \cite{Bishop, ChrucielIfsits,
VLSaw}.  The observation that the cosmological horizon is a Killing
horizon and hence an isolated horizon should be helpful in this regard.

\acknowledgments We would like to thank B\'{e}atrice Bonga for
discussions and clarifications regarding \cite{ABKIII}.

\appendix

\section{An averaging procedure}

In the main body we specified an averaging procedure by stipulating its
properties namely, (i) average of odd powers of $h$ vanishes and (ii)
average of space-time divergence is sub-leading. This was then used to
simplify the expression for the ripple stress tensor.
An averaging procedure satisfying these properties is indeed given by
Isaacson \cite{Isaacson}. We will use the same one and give more
explicit details in the present context.

Isaacson defines the {\em space-time average} of a tensor by using the
{\em parallel propagator bi-tensor, $g_{\mu}^{~\mu'}(x, x')$} as:
\begin{equation}\label{AverageDefn}
\langle X_{\mu\nu}\rangle (x) ~ := ~
\frac{\int_{cell}d^4x'\sqrt{|g(x')|}g_{\mu}^{~\mu'}(x,
x')g_{\nu}^{~\nu'}(x, x')X_{\mu'\nu'}(x')}{
\int_{cell}d^4x'\sqrt{|g(x')|}} \ .
\end{equation}

In the present context, we need average of the stress tensor for ripples
due to an retarded solution which has certain explicit form. We will use
this information to choose suitable integration variables and
corresponding `cell' denoting the averaging region. Because of this, we
have not used any weighting function as given by Isaacson
\cite{Isaacson}. 

To keep track of the powers of $H$, we begin by going from the conformal
chart ($\eta, x^i$) to the cosmological chart ($t, x^i$), $\eta := -
H^{-1}e^{-H t}$ with the spatial coordinates unchanged. In the
cosmological chart:
\begin{eqnarray} 
Metric & : & d s^2 ~ = ~ - dt^2 + a^2(t)(\delta_{ij}dx^i dx^j) ~~~,~~~
a(t) := e^{Ht} \\
Connection & : & \Gamma^t_{~tt} = 0 ~~,~~ \Gamma^t_{~tj} = 0 ~~,~~
\Gamma^t_{~ij} = Ha^2(t)\delta_{ij} \nonumber \\
& & \Gamma^i_{~tt} = 0 ~~,~~ \Gamma^i_{~tj} = H\delta^i_{~j} ~~,~~
\Gamma^i_{~jk} = 0 \ .
\end{eqnarray} 

The parallel propagator is computed in terms of the parallel transport
of an arbitrary co-tetrad (or tetrad): $g_{\mu}^{~\mu'}(x, x') :=
e_{\mu}^{~a}(x)e_a^{~\mu'}(x')$. The averaging region is small enough
that for a cell around a point $P$ with coordinates $x^{\alpha}$, there
is unique geodesic to points $P'$ with coordinates $x'^{\alpha}$. The
parallel transported co-tetrad is obtained using Taylor expansions of
the co-tetrad, the affine connection and the coordinates along the
geodesic, in terms of its affine parameter and eliminating the affine
parameter afterwards in favour of the coordinate differences $\Delta
x^{\alpha} := x'^{\alpha} - x^{\alpha}$. Details may be seen in the
appendix B of \cite{DateHoque}. There is a slight difference from
\cite{DateHoque} since that calculation was given in the context of
Fermi normal coordinates where the connection is already of order $H^2$
while in the cosmological chart, the connection is of order $H$. The
final expressions are:
\begin{eqnarray}
e_{\mu}^{~a}(x') & = &
\hat{e}_{\lambda}^{~a}\left[\delta_{\mu}^{~\lambda} +
\hat{\Gamma}_{\mu\alpha}^{~~\lambda}\Delta x^{\alpha} +
\frac{1}{2}\left(\widehat{\partial_{\rho}\Gamma}_{\mu\sigma}^{~~\lambda}
+
\hat{\Gamma}_{\mu\sigma}^{~~\alpha}\hat{\Gamma}_{\alpha\rho}^{~~\lambda}\right)\Delta
x^{\rho}\Delta x^{\sigma}\right] \\
g_{\mu}^{~\mu'}(x, x') & = & \delta_{\mu}^{~\mu'} -
\hat{\Gamma}_{\mu\alpha}^{~~\mu'}\Delta x^{\alpha} -
\frac{1}{2}\left(\widehat{\partial_{\rho}\Gamma}_{\mu\sigma}^{~~\mu'}
- 
\hat{\Gamma}_{\alpha\rho}^{~~\mu'}\hat{\Gamma}_{\sigma\mu}^{~~\alpha}\right)
\Delta x^{\rho}\Delta x^{\sigma}
\end{eqnarray}
In the above, the hatted quantities are evaluated at $x$. 

The connection dependent terms are linear and quadratic in $H\Delta x$.
Although the coordinate differences are much larger than the length
scale $\lambda$ they are much smaller than $H^{-1}$. Hence, these terms
can be neglected and {\em effectively the parallel propagator reduces to
just the Kronecker delta.} For purposes of illustration of averaging,
this suffices. It remains to integrate the $X_{\mu'\nu'}$ over the cell
and as noted in the main text in the paragraph below equation
(\ref{RhoFlux}), the components of the ripple stress tensor are
essentially determined in terms of $\partial_{\eta}\chi^{tt}_{ij} =
2\Case{\eta}{r} \Case{\mathcal{Q}^{tt}_{ij}(\eta -r)}{\eta - r}$ or
alternatively in terms of $\partial_{\eta}\chi_{ij}^{TT} =
2\Case{\eta}{r}\Case{\mathcal{R}^{TT}_{ij}(\eta - r)}{\eta - r}$. 

The angular dependence is introduced due to the `tt' part, eg as is
explicit in the $\Lambda_{ij}^{~~kl}(\hat{r})$ projector. The $(\eta,
r)$ dependence has a convenient factorised form. It is thus natural to
change the integration variables from $(\eta, r)$ to $(\bar{t}, \rho)$,
where $\bar{t}$ is the retarded synchronous time defined through, $\eta
-r := -H^{-1}e^{H\bar{t}}$ and $H\rho := -\Case{r}{\eta}$ defines
$\rho$. For definiteness, consider the average,
\begin{equation} \label{EtaEtaAve}
\langle\partial_{\eta}\chi_{mn}^{tt}\partial_{\eta}\chi_{tt}^{mn}\rangle
(t, r, \hat{r}) 
~ := ~ \frac{\int_{cell} dt\ dr\ r^2\ d^2s\ a^3(t)
\partial_{\eta}\chi_{mn}^{tt}(\bar{t}) \partial_{\eta}\chi^{mn}_{tt}
(\bar{t})} { \int_{cell}dt\ dr\ r^2\ d^2s\ a^3(t)}
\end{equation}
Here $\hat{r}$ denotes a point on $S^2$ (a spatial direction). We will
specify the cell after changing over to $(\bar{t}, \rho, \hat{r})$.

From the definitions, we arrive at the coordinate transformations,
\begin{eqnarray}
\eta(\bar{t}, \rho) & = & - \frac{1}{H(1 + H\rho)}e^{-H\bar{t}} ~~ , ~~
r(\bar{t}, \rho) ~=~ \frac{\rho}{1 + H\rho}e^{-H\bar{t}} ~~ \Rightarrow
\nonumber \\
a(t) & = & a(\bar{t})(1 + H\rho) \hspace{1.2cm}, ~~\mbox{with} ~~~ a(t)
:= e^{H t} \ . 
\end{eqnarray}

The Jacobian of transformation is $\Case{\partial(t,
r)}{\partial(\bar{t}, \rho)} = \{a(\bar{t})(1 + H\rho)\}^{-1}$.  We
choose the cell so that $\bar{t} \in [\bar{t}_0 - \delta, \bar{t}_0 +
\delta]$ and $\rho \in [\rho_0 - \Delta, \rho_0 + \Delta]$ and $\hat{r}
\in \Delta\omega$.  The coordinate windows $\delta, \Delta$ and
$\sqrt{r^2\Delta\omega}$ are several times the ripple scale while
$(\bar{t}_0, \rho_0)$ are the transforms of $(t, r)$. In terms of these
choices, the average becomes,
\begin{eqnarray}\label{EtaEtaAverage}
\langle\partial_{\eta}\chi_{mn}^{tt}\partial_{\eta}\chi_{tt}^{mn}\rangle
(t, r, \hat{r}) 
& := & \frac{\int_{\bar{t}_0 -~\delta}^{\bar{t}_0 + \delta} d\bar{t}\
\int_{\rho_0 - \Delta}^{\rho_0 + \Delta}d\rho\ \rho^2
\int_{\Delta\omega}d^2s\
\left[ 4\frac{a^2(\bar{t})}{\rho^2}\mathcal{Q}_{mn}^{tt}(\bar{t})
\mathcal{Q}^{mn}_{tt}(\bar{t})\right]}
{\int_{\bar{t}_0 - \delta}^{\bar{t}_0 + \delta} d\bar{t}\ \int_{\rho_0 -
\Delta}^{\rho_0 + \Delta}d\rho\ \rho^2 \int_{\Delta\omega}d^2s\ }
\end{eqnarray}

Consider the angular integration. The angular dependence arises in
taking the `tt' part of the solution $\chi_{ij}(\eta, r)$. For
illustration purpose, consider $r$ to be sufficiently large so that we
can use the $\Lambda_{ij}^{~~kl}(\hat{r})$ projector, giving
$\partial_{\eta}\chi_{ij}^{tt}\partial_{\eta}\chi^{ij}_{tt} \sim
\Lambda_{ij}^{~~kl}\partial_{\eta}\chi^{ij}\partial_{\eta}\chi_{kl}$.
For large $r$, the angular coordinate windows are $\sim \lambda/r \ll
1$. Using the mean value theorem in the angular integration in the
numerator, we get 
\begin{equation}\label{AngularAverage}
\frac{\int_{\Delta\omega}d^2s(\hat{r}')\Lambda_{ij}^{~~kl}(\hat{r}')}
{\int_{\Delta\omega}d^2s(\hat{r}')} ~\approx~
\Lambda_{ij}^{~~kl}(\hat{r}) \ .
\end{equation}
In effect, the $\Lambda-$projector comes out of the averaging and the
{{\em angular average trivializes}}. Of the remaining integrations,
the $\rho$ integration can be done explicitly and is independent of
$\Delta$ to the leading order in $\Delta/\rho_0$.  Thus, in the
numerator of (\ref{EtaEtaAverage}) we get,
\begin{eqnarray}
\int_{\bar{t}_0 - \delta}^{\bar{t}_0 + \delta}d\bar{t}\int_{\rho_0 -
\Delta}^{\rho_0 + \Delta}d\rho~~ 4\
a^2(\bar{t})\mathcal{Q}^{ij}(\bar{t})\ \mathcal{Q}_{kl}(\bar{t})
& \approx & 8\Delta \int_{\bar{t}_0 - \delta}^{\bar{t}_0 +
\delta}d\bar{t}~~ a^2(\bar{t})\mathcal{Q}^{ij}(\bar{t})\
\mathcal{Q}_{kl}(\bar{t}) \nonumber \\
& = & 8\Delta\int_{-\delta}^{\delta}dy\ a^2(\bar{t}_0 + y)
\mathcal{Q}^{ij} \mathcal{Q}_{kl}(\bar{t}_0 + y) \nonumber \\
& = & 8\Delta a^2(\bar{t}_0)\int_{-\delta}^{\delta}dy\ a^2(y)
\mathcal{Q}^{ij} \mathcal{Q}_{kl}(\bar{t}_0 + y) \nonumber \\
& := & (8\Delta)a^2(\bar{t}_0) (2\delta)\langle\mathcal{Q}^{ij}
\mathcal{Q}_{kl}\rangle_{\bar{t}}(\bar{t}_0) \label{Numerator} 
\end{eqnarray}
In the third line, we have used $a^2$ being an exponential function and
in the last line we have {\em defined} the average over the retarded
time around $\bar{t}_0$ and put the suffix on the angular bracket
as a reminder. 

The $\langle\rangle_{\bar{t}}$ averaging has the extra factor of
$a^2(y)$. However, over the integration domain $(-\delta, \delta)$, we
can approximate $a^2(y) \approx 1 + 2Hy + \cdots$ and neglect $o(Hy)$
terms since $H\delta \sim k\lambda/L \sim k\epsilon \ll 1$. The extra
factor thus introduces a small deviation from the usual averaging
without the extra factor and {\em we neglect it henceforth and the
reminder suffix, $\bar{t}$ is also suppressed.}

In the denominator we get,
\begin{eqnarray}
\int_{\bar{t}_0 - \delta}^{\bar{t}_0 + \delta}d\bar{t}\int_{\rho_0 -
\Delta}^{\rho_0 + \Delta}d\rho~ \rho^2 
& \approx & 2\rho^2_0\Delta \int_{\bar{t}_0 - \delta}^{\bar{t}_0 +
\delta}d\bar{t} ~ = ~ (2\Delta)(2\delta)\rho^2_0 \label{Denominator} 
\end{eqnarray}
Combining equations (\ref{EtaEtaAverage}, \ref{AngularAverage},
\ref{Numerator}, \ref{Denominator}), we get
\begin{eqnarray} \label{TEtaEta}
\langle\partial_{\eta}\chi_{mn}^{tt}\partial_{\eta}\chi_{tt}^{mn}\rangle
(t, r, \hat{r}) 
& = & 4\frac{a^2(\bar{t}_0)}{\rho_0^2} \langle \mathcal{Q}_{ij}^{tt}\
\mathcal{Q}^{ij}_{tt}\rangle (\bar{t}_0, \hat{r})
\end{eqnarray}
In the last equation, we have combined the averaging over retarded time
and the (trivial) angular average. We have also inserted the
$\Lambda-$projector.  {The averaging over {\em a space-time cell}
has been reduced to averaging over {\em a 3-dimensional cell on a $\rho
= $constant hypersurface}}. The pre-factor on the right hand side of the
above equation exactly equals the last square bracket in the first line
of the equation (\ref{RhoFlux}). In effect, the $(\Case{2\eta}{r(\eta -
r)})^2$ factor has come out of the averaging.

{ We can also reduce the space-time average to a hypersurface
average for
$\partial_{\eta}\chi_{ij}^{TT}\partial_{\eta}\chi^{ij}_{TT}$. Following
the same steps as from eqn.(\ref{EtaEtaAve}) onwards, we will arrive at
eqn.(\ref{EtaEtaAverage}) with $\mathcal{Q}^{tt}_{mn} \to
\mathcal{Q}^{TT}_{mn}$. We cannot do the angular averaging as before,
but we don't need to. Crucially, the $\rho$ dependence has factored out
exactly as before and the average over $\rho$ gives $\rho_0^{-2}$ as
before. The $\bar{t}$ averaging too gives $a^2(\bar{t}_0)$ and we get
the desired result,
\begin{eqnarray} \label{TTEtaEta}
\langle\partial_{\eta}\chi_{mn}^{TT}\partial_{\eta}\chi_{TT}^{mn}\rangle
(t, r, \hat{r}) 
& = & 4\frac{a^2(\bar{t}_0)}{\rho_0^2} \langle \mathcal{Q}_{ij}^{TT}\
\mathcal{Q}^{ij}_{TT}\rangle (\bar{t}_0, \hat{r})
\end{eqnarray}

}

We can relate the averaging over the retarded time, $\bar{t}$, to the
averaging over the Killing time, $\tau$ along the $\rho = \rho_0$ curve.
From the coordinate transformation, we have $\eta - r = -
H^{-1}e^{-H\bar{t}}$ while along $\rho = \rho_0$ Killing trajectory,
$\eta - r = (\eta_* - r_*)e^{-H\tau} = - (H^{-1}e^{-H\bar{t}_*})
e^{-H\tau}$. Hence, $\bar{t} = \tau + \bar{t}_*$ and the temporal
averaging is related to averaging over a Killing time. {Note that
the averaging cell being bounded by two hypersurfaces of constant
retarded times, the temporal averaging may be evaluated along the source
worldline, $r = 0$ {\em or} along the Killing trajectory on
$\mathcal{J}^+$.}

We also have mixed and spatial components of the ripple stress tensor.
These involve $\partial_i\chi_{mn}^{tt} \approx
\hat{x}_i\partial_r\chi_{mn}^{tt} \approx - \hat{x}_i\partial_{\eta}
\chi_{mn}^{tt}$. While taking the average, the $\hat{x}_i$ can be taken
out of the average since the angular coordinate windows are of very
small size $\sim \lambda/r$. This allows us to take $\hat{x}^i$ across
the angular averages and replace all components of the ripple stress
tensor by $t_{\eta\eta}$ in the conformal chart or by $t_{00}$ in the
cosmological chart.


\end{document}